\documentclass[thmsa]{article}
\usepackage{sw20lart}

\input tcilatex
\QQQ{Language}{
British English
}

\begin{document}

\section{\bf Introduction}

We have recently shown how spin-$1/2$ vectors and matrices can be derived
from first principles$[1,2]$. The method used is based on the Land\'e
interpretation of quantum mechanics $[3,6]$. This essentially consists in
using the probability amplitudes which characterize spin measurements
between different directions as the basis set for the expansion of the same
probability amplitudes. The method is valid for all values of spin, and we
have illustrated it by applying it to the case of spin $1/2$. We have
thereby obtained not only the Pauli spin matrices and vectors but the most
general forms of these quantities as well.

The subject of this paper is the derivation of the spin vectors and
operators for the case of spin $1$. This task is far from routine because to
the best of our knowledge, the generalized results we shall introduce are
new. At the same time, we confirm the validity of the method. The fact that
it works for both spin $1/2$ and spin $1$ indicates that it will apply to
any value of $J$. Our success in this program shows the credibility of the
Land\'e interpretation of quantum mechanics.

In this paper, we demonstrate how to obtain the standard forms of the spin -$%
1$ operators and eigenvectors from first principles, and how to generalize
their forms. After deriving the standard general formulas for the
probability amplitudes that describe spin-$1$ measurements, we use them to
find the most generalized forms of the operators for the $x$, $y$ and $z$
components of spin and their eigenvectors. We also obtain the generalized
ladder operators. The correctness of these quantities is proved by the fact
that in the appropriate limit, they yield the standard forms. The operators
for the $x$, $y$ and $z$ components satisfy the commutation relations, and
the generalized ladder operators have the required effect on the
eigenvectors of the generalized $z$ component of spin. In addition to these
operators, we derive the operator for the square of the spin. As a final
step, we summarize the procedure to be used in applying our method to any
value of $J$.

Our method is justified fully and developed step by step in ref. $1$. Here
we shall use the method with a minimum of the underlying theory.

This paper is structured as follows. In Section $2$, we review those results
of the Land\'e approach which we shall need to use. Here, we give the
general form of the matrix eigenvalue equation in this approach, as well as
the form of the matrix operator and its eigenvectors. In Section $3$, we
apply these results to the case of spin $1$. Thus, in Subsection $3.1$, we
derive the forms of the operators and vectors for spin $1$ in terms of
implicit probability amplitudes for spin projection measurements. In
Subsection $3.2$, we summarize all the possible forms that the operators and
vectors can take if we vary some of the parameters on which they depend.

We show in Subsection $3.3$ that we are able to derive the standard forms of
the operators and vectors without explicit knowledge of the probability
amplitudes. We then derive the explicit forms of the probability amplitudes
in Subsection $3.4$, and verify their correctness in Subsection $3.5$ by
showing that they obey the Land\'e expansion for probability amplitudes,
which the whole theory is based on. We obtain the corresponding
probabilities from them in Subsection $3.6$. Armed with the explicit
probability amplitudes, we obtain the generalized form of the operator for $%
z $ component of the spin as well as its eigenvectors in Subsection $3.7$.

In Subsection $3.8$, we give the generalized forms of the $x$ and $y$
components of the spin and their eigenvectors. We observe that the three
generalized components obey the usual commutation relations. In Subsection $%
3.9$ we give the ladder operators, which are found to perform their lowering
and raising functions on the generalized eigenvectors of the $z$ component
of the spin operator. We derive the operator for the square of the spin in
Subsection $3.10$. In Subsection $3.11$, we list some properties of the
generalized probability amplitudes.

In Section $4$, we summarize the steps to follow in order to apply this
method to higher values of $J$. We conclude the paper after a discussion in
Section $5$.

\section{\bf Basic Results}

Let three observables $A$, $B$ and $C$ belong to the same quantum system.
Let each one have $N$ eigenvalues. If the system is known to be in a state
corresponding to the eigenvalue $A_i$ of $A$ when $B$ is measured, then the
probability amplitudes corresponding to the different values of $B$ being
obtained are $\chi (A_i;B_k)$ where $k=1,2,...,N.$ But if $C$ is measured,
the probability amplitudes corresponding to the different values of $C$
being obtained are $\psi (A_i;C_k)$ where $k=1,2,...,N.$ Finally, if $C$ is
measured after the system has been ascertained to be in a state
corresponding to $B_i$ the probability amplitudes for obtaining the
different values of $C$ are $\phi (B_i;C_k)$ where $k=1,2,...,N.$

The probability amplitudes obey two-way symmetry: hence they satisfy the
Hermiticity condition

\begin{equation}
\psi (A_i;C_k)=\psi ^{*}(C_k;A_i).  \label{one}
\end{equation}

They are related by the Land\'e expansion [3-6]

\begin{equation}
\psi (A_i;C_k)=\dsum\limits_{j=1}^N\chi (A_i;B_j)\phi (B_j;C_k).  \label{two}
\end{equation}

Using this basic formula, we find that the matrix representation of the
probability amplitude $\psi (A_i;C_k)$ in the basis $\phi (B_j;C_k)$ is [$1$]

\begin{equation}
\lbrack \psi (A_i;C_k)]=\left( 
\begin{array}{c}
\chi (A_i;B_1) \\ 
\chi (A_i;B_2) \\ 
... \\ 
\chi (A_i;B_N)
\end{array}
\right) .  \label{three}
\end{equation}
This formula is essentially the one found in standard texts. It differs in
giving the elements of the vector as probability amplitudes connecting an
initial state corresponding to eigenvalue $A_i$ with a set of intermediate
states corresponding to another observable - in this case $B$.

Consider now the differential eigenvalue equation

\begin{equation}
A(x)\psi (\lambda _k;x)=\lambda _k\psi (\lambda _k;x)  \label{four}
\end{equation}
where $\lambda _k$ is an eigenvalue, and $x$ is a general quantity which
represents the final eigenvalue, and whose spectrum is continuous. This
eigenvalue equation emphasizes the idea that the eigenfunction is a
probability amplitude connecting an initial state corresponding to the
eigenvalue $\lambda _k$ with a set of final states whose eigenvalues are
given by $x$. We use Eq. (\ref{two}) to express $\psi (\lambda _k;x)\;$as an
expansion:

\begin{equation}
\psi (\lambda _k;x)=\dsum\limits_{j=1}^N\chi (\lambda _k;B_j)\phi (B_j;x).
\label{six}
\end{equation}
Substituting this in the eigenvalue equation Eq. (\ref{four}), we obtain [$1$%
] the matrix eigenvalue equation which is the equivalent of the differential
eigenvalue equation :

\begin{equation}
\left( 
\begin{array}{cccc}
A_{11}-\lambda _k & A_{12} & ... & A_{1N} \\ 
A_{21} & A_{22}-\lambda _k & ... & A_{2N} \\ 
... & ... & ... & ... \\ 
... & ... & ... & ... \\ 
A_{N1} & A_{N2} & ... & A_{NN}-\lambda _k
\end{array}
\right) \left( 
\begin{array}{c}
\chi (\lambda _k;B_1) \\ 
\chi (\lambda _k;B_2) \\ 
... \\ 
... \\ 
\chi (\lambda _k;B_N)
\end{array}
\right) =0,  \label{seven}
\end{equation}
where 
\begin{equation}
A_{mj}=\left\langle \phi (B_m;x)\left| A(x)\right| \phi (B_j;x)\right\rangle
.  \label{eight}
\end{equation}
The expectation value of the quantity $R(x)$ is given by the formula

\begin{equation}
\left\langle R\right\rangle =\dsum\limits_{i=1}^N\dsum\limits_{j=1}^N\chi
^{*}(\lambda _k;B_i)R_{ij}\chi (\lambda _k;B_j)=[\psi _k]^{\dagger }\left[
R\right] [\psi \chi _k],  \label{nine}
\end{equation}
where

\begin{equation}
\lbrack \psi _k]=\left( 
\begin{array}{c}
\chi (\lambda _k;B_1) \\ 
\chi (\lambda _k;B_2) \\ 
... \\ 
... \\ 
\chi (\lambda _k;B_N)
\end{array}
\right)  \label{ten}
\end{equation}
and

\begin{equation}
\left[ R\right] =\left( 
\begin{array}{cccc}
R_{11} & R_{12} & ... & R_{1N} \\ 
R_{21} & R_{22} & ... & R_{2N} \\ 
... & ... & ... & ... \\ 
... & ... & ... & ... \\ 
R_{N1} & R_{N2} & ... & R_{NN}
\end{array}
\right) .  \label{el11}
\end{equation}

The matrix elements of $\left[ R\right] $ are 
\begin{equation}
R_{ij}=\left\langle \phi _i\left| R(x)\right| \phi _j\right\rangle .
\label{tw12}
\end{equation}

\section{\bf Specialization To Spin 1 Systems}

\subsection{\bf Derivation of General Matrix Mechanics Formulas}

The Land\'e formula Eq. (\ref{two}) applies to any three sets of probability
amplitudes which relate different observables of a quantum system. As we
showed in ref. $1$, the formula is particularly useful for the treatment of
spin. The reason is that spin projections with respect to different
reference vectors have the status of different observables. Yet, the three
sets of probability amplitudes have the same functional form. This allows us
to deduce the probability amplitudes easily.

Consider a system whose spin is known to be up with respect to the unit
vector $\widehat{{\bf a}}$. We are interested in the probability amplitudes
for measurement of the spin along the new direction determined by the unit
vector $\widehat{{\bf c}}{\bf .}$ We are interested in the quantity $R({\bf %
\sigma }\cdot \widehat{{\bf c}})$, which is a function of the spin
projection ${\bf \sigma }\cdot \widehat{{\bf c}}.$ We may also measure the
spin projection with respect to the unit vector $\widehat{{\bf b}}$, and the
value of the $R({\bf \sigma }\cdot \widehat{{\bf b}})$.

We shall measure the spin in units of $\hbar .$ Thus, the possible values of
the spin projection are $+1$, $0$ and $-1$. The probability amplitudes will
be labelled with the magnetic quantum numbers corresponding to the initial
and final directions; the quantization direction for each quantum number
will be given as a subscript of that quantum number. Thus for measurements
from the initial direction $\widehat{{\bf a}}$ to the final direction $%
\widehat{{\bf c}}$, the complete set of possible probability amplitudes is $%
\psi (m_i^{(\widehat{{\bf a}})};m_f^{(\widehat{{\bf c}})})$, with all
possible nine combinations $m_i^{(\widehat{{\bf a}})},m_f^{(\widehat{{\bf c}}%
)}=-1,0,1.$ $\;$

For measurement from the initial direction $\widehat{{\bf a}}$ to the final
direction $\widehat{{\bf b}}$, the probability amplitudes are $\chi (m_i^{(%
\widehat{{\bf a}})};m_f^{(\widehat{{\bf b}})})$, again with $m_i^{(\widehat{%
{\bf a}})},m_f^{(\widehat{{\bf b}})}=-1,0,1.$

For measurement from the initial direction $\widehat{{\bf b}}$ to the final
direction $\widehat{{\bf c}}$, the probability amplitudes are $\phi (m_i^{(%
\widehat{{\bf b}})};m_f^{(\widehat{{\bf c}})}).$ In all cases, the ordering
of the eigenvalues is such that $i,f=1$ corresponds to the eigenvalue $+1$, $%
i,f=2$ corresponds to the eigenvalue $0$ and $i,f=3$ to the eigenvalue $-1$.

According to the Land\'e formula, Eq. (\ref{two}), the following expansions
hold for the probability amplitudes $\psi :$

\begin{eqnarray}
\psi (m_i^{(\widehat{{\bf a}})};m_f^{(\widehat{{\bf c}})}) &=&\chi (m_i^{(%
\widehat{{\bf a}})};(+1)^{(\widehat{{\bf b}})})\phi ((+1)^{(\widehat{{\bf b}}%
)};m_f^{(\widehat{{\bf c}})})  \nonumber \\
&&+\chi (m_i^{(\widehat{{\bf a}})};0^{(\widehat{{\bf b}})})\phi (0^{(%
\widehat{{\bf b}})};m_f^{(\widehat{{\bf c}})})  \nonumber \\
&&+\chi (m_i^{(\widehat{{\bf a}})};(-1)^{(\widehat{{\bf b}})})\phi ((-1)^{(%
\widehat{{\bf b}})};m_f^{(\widehat{{\bf c}})}),  \label{th13}
\end{eqnarray}

Similar expressions hold for the probability amplitudes $\chi $ and $\phi .$
The matrix representations of the probability amplitudes are 
\begin{equation}
\lbrack \psi (m_i^{(\widehat{{\bf a}})};m_f^{(\widehat{{\bf c}})})]=\left( 
\begin{array}{c}
\chi (m_i^{(\widehat{{\bf a}})};(+1)^{(\widehat{{\bf b}})}) \\ 
\chi (m_i^{(\widehat{{\bf a}})};0^{(\widehat{{\bf b}})}) \\ 
\chi (m_i^{(\widehat{{\bf a}})};(-1)^{(\widehat{{\bf b}})})
\end{array}
\right)  \label{tw22}
\end{equation}

Let the quantity $R({\bf \sigma }\cdot \widehat{{\bf c}})$ have the value $%
r_1$ when the spin projection is $+1$, $r_2\;$when the spin projection is $0$%
, and $r_3$ when the spin projection is $-1$. Assuming that the initial
state corresponds to the spin projection $m_i^{(\widehat{{\bf a}})}\hbar ,$
the expectation value of $R$ is

\begin{equation}
\left\langle R\right\rangle =\sum_{n=1}^3\left| \psi (m_i^{(\widehat{{\bf a}}%
)};m_n^{(\widehat{{\bf c}})})\right| ^2r_n  \label{tw25}
\end{equation}

In consequence of the expansions given by Eqn. (\ref{th13}), the expectation
value Eq. (\ref{tw25}) may be written as

\begin{equation}
\left\langle R\right\rangle =[\psi (m_i^{(\widehat{{\bf a}})})]^{\dagger
}\left[ R\right] [\psi (m_i^{(\widehat{{\bf a}})})],  \label{tw26}
\end{equation}
where

\begin{equation}
\lbrack \psi (m_i^{(\widehat{{\bf a}})})]=\left( 
\begin{array}{c}
\chi (m_i^{(\widehat{{\bf a}})};(+1)^{(\widehat{{\bf b}})}) \\ 
\chi (m_i^{(\widehat{{\bf a}})};0^{(\widehat{{\bf b}})}) \\ 
\chi (m_i^{(\widehat{{\bf a}})};(-1)^{(\widehat{{\bf b}})})
\end{array}
\right)  \label{tw27}
\end{equation}
and

\begin{equation}
\left[ R\right] =\left( 
\begin{array}{ccc}
R_{11} & R_{12} & R_{13} \\ 
R_{21} & R_{22} & R_{23} \\ 
R_{31} & R_{32} & R_{33}
\end{array}
\right) .  \label{tw28}
\end{equation}
The elements of $[R]$ are

\begin{eqnarray}
R_{11} &=&\left| \phi ((+1)^{(\widehat{{\bf b}})};(+1)^{(\widehat{{\bf c}}%
)})\right| ^2r_1+\left| \phi ((+1)^{(\widehat{{\bf b}})};0^{(\widehat{{\bf c}%
})})\right| ^2r_2  \nonumber \\
&&+\left| \phi ((+1)^{(\widehat{{\bf b}})};(-1)^{(\widehat{{\bf c}}%
)})\right| ^2r_3,  \label{tw29}
\end{eqnarray}

\begin{eqnarray}
R_{12} &=&\phi ^{*}((+1)^{(\widehat{{\bf b}})};(+1)^{(\widehat{{\bf c}}%
)})\phi (0^{(\widehat{{\bf b}})};(+1)^{(\widehat{{\bf c}})})r_1++\phi
^{*}((+1)^{(\widehat{{\bf b}})};0^{(\widehat{{\bf c}})})\phi (0^{(\widehat{%
{\bf b}})};0)^{(\widehat{{\bf c}})})r_2  \nonumber \\
&&+\phi ^{*}((+1)^{(\widehat{{\bf b}})};(-1)^{(\widehat{{\bf c}})})\phi (0^{(%
\widehat{{\bf b}})};(-1)^{(\widehat{{\bf c}})})r_3,  \label{th30}
\end{eqnarray}
\begin{eqnarray}
R_{13} &=&\phi ^{*}((+1)^{(\widehat{{\bf b}})};(+1)^{(\widehat{{\bf c}}%
)})\phi ((-1)^{(\widehat{{\bf b}})};(+1)^{(\widehat{{\bf c}})})r_1+\phi
^{*}((+1)^{(\widehat{{\bf b}})};0^{(\widehat{{\bf c}})})\phi ((-1)^{(%
\widehat{{\bf b}})};0^{(\widehat{{\bf c}})})r_2  \nonumber \\
&&+\phi ^{*}((+1)^{(\widehat{{\bf b}})};(-1)^{(\widehat{{\bf c}})})\phi
((-1)^{(\widehat{{\bf b}})};(-1)^{(\widehat{{\bf c}})})r_3,  \label{th31}
\end{eqnarray}
\begin{eqnarray}
R_{21} &=&\phi ^{*}(0^{(\widehat{{\bf b}})};(+1)^{(\widehat{{\bf c}})})\phi
((+1)^{(\widehat{{\bf b}})};(+1)^{(\widehat{{\bf c}})})r_1+\phi ^{*}(0^{(%
\widehat{{\bf b}})};0^{(\widehat{{\bf c}})})\phi ((+1)^{(\widehat{{\bf b}}%
)};0^{(\widehat{{\bf c}})})r_2  \nonumber \\
&&+\phi ^{*}(0^{(\widehat{{\bf b}})};(-1)^{(\widehat{{\bf c}})})\phi ((+1)^{(%
\widehat{{\bf b}})};(-1)^{(\widehat{{\bf c}})})r_3,  \label{th32}
\end{eqnarray}
\begin{equation}
R_{22}=\left| \phi (0^{(\widehat{{\bf b}})};(+1)^{(\widehat{{\bf c}}%
)})\right| ^2r_1+\left| \phi (0^{(\widehat{{\bf b}})};0^{(\widehat{{\bf c}}%
)})\right| ^2r_2+\left| \phi (0^{(\widehat{{\bf b}})};(-1)^{(\widehat{{\bf c}%
})})\right| ^2r_3,  \label{th33}
\end{equation}
\begin{eqnarray}
R_{23} &=&\phi ^{*}(0^{(\widehat{{\bf b}})};(+1)^{(\widehat{{\bf c}})})\phi
((-1)^{(\widehat{{\bf b}})};(+1)^{(\widehat{{\bf c}})})r_1+\phi ^{*}(0^{(%
\widehat{{\bf b}})};0^{(\widehat{{\bf c}})})\phi ((-1)^{(\widehat{{\bf b}}%
)};0^{(\widehat{{\bf c}})})r_2  \nonumber \\
&&+\phi ^{*}(0^{(\widehat{{\bf b}})};(-1)^{(\widehat{{\bf c}})})\phi ((-1)^{(%
\widehat{{\bf b}})};(-1)^{(\widehat{{\bf c}})})r_3,  \label{th34}
\end{eqnarray}
\begin{eqnarray}
R_{31} &=&\phi ^{*}((-1)^{(\widehat{{\bf b}})};(+1)^{(\widehat{{\bf c}}%
)})\phi ((+1)^{(\widehat{{\bf b}})};(+1)^{(\widehat{{\bf c}})})r_1+\phi
^{*}((-1)^{(\widehat{{\bf b}})};0^{(\widehat{{\bf c}})})\phi ((+1)^{(%
\widehat{{\bf b}})};0^{(\widehat{{\bf c}})})r_2  \nonumber \\
&&+\phi ^{*}((-1)^{(\widehat{{\bf b}})};(-1)^{(\widehat{{\bf c}})})\phi
((+1)^{(\widehat{{\bf b}})};(-1)^{(\widehat{{\bf c}})})r_3,  \label{th35}
\end{eqnarray}
\begin{eqnarray}
R_{32} &=&\phi ^{*}((-1)^{(\widehat{{\bf b}})};(+1)^{(\widehat{{\bf c}}%
)})\phi (0^{(\widehat{{\bf b}})};(+1)^{(\widehat{{\bf c}})})r_1+\phi
^{*}((-1)^{(\widehat{{\bf b}})};0^{(\widehat{{\bf c}})})\phi (0^{(\widehat{%
{\bf b}})};0^{(\widehat{{\bf c}})})r_2  \nonumber \\
&&+\phi ^{*}((-1)^{(\widehat{{\bf b}})};(-1)^{(\widehat{{\bf c}})})\phi (0^{(%
\widehat{{\bf b}})};(-1)^{(\widehat{{\bf c}})})r_3  \label{th36}
\end{eqnarray}
and 
\begin{equation}
R_{33}=\left| \phi ((-1)^{(\widehat{{\bf b}})};(+1)^{(\widehat{{\bf c}}%
)})\right| ^2r_1+\left| \phi ((-1)^{(\widehat{{\bf b}})};0^{(\widehat{{\bf c}%
})})\right| ^2r_2+\left| \phi ((-1)^{(\widehat{{\bf b}})};(-1)^{(\widehat{%
{\bf c}})})\right| ^2r_3.  \label{th37}
\end{equation}

These matrix elements are given by

\begin{equation}
R_{kl}=\sum \phi ^{*}(m_k^{(\widehat{{\bf b}})};m_n^{(\widehat{{\bf c}}%
)})\phi (m_l^{(\widehat{{\bf b}})};m_n^{(\widehat{{\bf c}})})r_n.
\label{th37z}
\end{equation}

\subsection{\bf Summary of the Various Possible Choices for the Reference
Vectors}

The forms that the quantities $[\psi ]$ and $\left[ R\right] $ assume are
determined by the choices we make for the vectors $\widehat{{\bf b}}$ and $%
\widehat{{\bf c}}$. The various possibilities are listed below.

{\bf Case (a)}: $\widehat{{\bf b}}\neq \widehat{{\bf a}}$ and $\widehat{{\bf %
c}}\neq \widehat{{\bf a}}$.

This is the most general case. In this case, the matrix representations are 
\begin{equation}
\lbrack \psi (m_i^{(\widehat{{\bf a}})};m_f^{(\widehat{{\bf c}})})]=\left( 
\begin{array}{c}
\chi (m_i^{(\widehat{{\bf a}})};(+1)^{(\widehat{{\bf b}})}) \\ 
\chi (m_i^{(\widehat{{\bf a}})};0^{(\widehat{{\bf b}})}) \\ 
\chi (m_i^{(\widehat{{\bf a}})};(-1)^{(\widehat{{\bf b}})})
\end{array}
\right)  \label{th38}
\end{equation}
while the elements of the operator $\left[ R\right] $ are given by Eqns. (%
\ref{tw29}) - (\ref{th37}).

{\bf Case (b)}: $\widehat{{\bf b}}=\widehat{{\bf a}}.$

The matrix representations are 
\begin{equation}
\lbrack \psi (m_i^{(\widehat{{\bf a}})};m_f^{(\widehat{{\bf c}})})]=\left( 
\begin{array}{c}
\delta _{m_i1} \\ 
\delta _{m_i2} \\ 
\delta _{m_i3}
\end{array}
\right) ,  \label{fo41}
\end{equation}
while the elements of $\left[ R\right] $ are given by Eqns. (\ref{tw29}) - (%
\ref{th37}) with $\widehat{{\bf b}}$ replaced by $\widehat{{\bf a}}$.

{\bf Case (c)}: $\widehat{{\bf b}}=\widehat{{\bf c}}.$

For this case, the matrix representations are 
\begin{equation}
\lbrack \psi (m_i^{(\widehat{{\bf a}})};m_f^{(\widehat{{\bf c}})})]=\left( 
\begin{array}{c}
\chi (m_i^{(\widehat{{\bf a}})};(+1)^{(\widehat{{\bf c}})}) \\ 
\chi (m_i^{(\widehat{{\bf a}})};0^{(\widehat{{\bf c}})}) \\ 
\chi (m_i^{(\widehat{{\bf a}})};(-1)^{(\widehat{{\bf c}})})
\end{array}
\right)  \label{fi53}
\end{equation}
while the operator is

\begin{equation}
\left[ R\right] =\left( 
\begin{array}{ccc}
r_1 & 0 & 0 \\ 
0 & r_2 & 0 \\ 
0 & 0 & r_3
\end{array}
\right) .  \label{fi56}
\end{equation}

{\bf Case (d)}: $\widehat{{\bf c}}=\widehat{{\bf a}}$.

The matrix representations are

\begin{equation}
\lbrack \psi (m_i^{(\widehat{{\bf a}})};m_f^{(\widehat{{\bf a}})})]=\left( 
\begin{array}{c}
\chi (m_i^{(\widehat{{\bf a}})};(+1)^{(\widehat{{\bf b}})}) \\ 
\chi (m_i^{(\widehat{{\bf a}})};0^{(\widehat{{\bf b}})}) \\ 
\chi (m_i^{(\widehat{{\bf a}})};(-1)^{(\widehat{{\bf b}})})
\end{array}
\right)  \label{fi57}
\end{equation}
while the elements of $\left[ R\right] $ are given by Eqns. (\ref{tw29}) - (%
\ref{th37}) with $\widehat{{\bf c}}$ replaced by $\widehat{{\bf a}}$.

{\bf Case (e}): $\widehat{{\bf b}}=\widehat{{\bf a}}$ and $\widehat{{\bf c}}=%
\widehat{{\bf a}}$.

The matrix representations are

\begin{equation}
\lbrack \psi (m_i^{(\widehat{{\bf a}})};m_f^{(\widehat{{\bf a}})})]=\left( 
\begin{array}{c}
\delta _{m_i1} \\ 
\delta _{m_i2} \\ 
\delta _{m_i3}
\end{array}
\right) .  \label{si60}
\end{equation}
while the operator is

\begin{equation}
\left[ R\right] =\left( 
\begin{array}{ccc}
r_1 & 0 & 0 \\ 
0 & r_2 & 0 \\ 
0 & 0 & r_3
\end{array}
\right) .  \label{si63}
\end{equation}

This case evidently corresponds to the standard forms of both the vectors
and the operator, and the vector states happen to be the eigenvectors of the
operator.

\subsection{\bf Derivation of the Standard Forms of the Spin Vectors and
Matrices }

We now employ the expressions for the elements of $[R]$ to deduce the matrix
forms of the spin operators. When $R$ is the spin projection itself, then $%
r_1=1,$ $r_2=0$ and $r_3=-1.\;$Cases (c) and (e) permit us to deduce the
standard form of the operator for the spin projection along the axis of
quantization. We shall call this the operator for the ''$z$ component'' of
spin, the quantization direction is not necessarily the $z$ direction. We
find that

\begin{equation}
\left[ \sigma _z\right] =\left( 
\begin{array}{ccc}
1 & 0 & 0 \\ 
0 & 0 & 0 \\ 
0 & 0 & -1
\end{array}
\right) .  \label{si64}
\end{equation}

We emphasize that the special form of \ ''$\left[ \sigma _z\right] "$ in Eq.
(\ref{si64}) applies whenever we choose $\widehat{{\bf b}}=\widehat{{\bf c}}$%
, even if neither $\widehat{{\bf a}}$ nor $\widehat{{\bf b}}$ nor $\widehat{%
{\bf c}}$ coincides with the $z$ direction. Nevertheless we find it natural
to give this operator the subscript $z$.

Now, since the eigenvalue of $\sigma ^2$ is $2$, we have $r_1=r_2=r_3=2$:
the standard form of the operator for the square of the spin is given by
Cases (c) and (e) as

\begin{equation}
\left[ \sigma ^2\right] =\left( 
\begin{array}{ccc}
2 & 0 & 0 \\ 
0 & 2 & 0 \\ 
0 & 0 & 2
\end{array}
\right) .  \label{si65}
\end{equation}

For the two cases $\widehat{{\bf b}}=\widehat{{\bf c}}$ and $\widehat{{\bf a}%
}=\widehat{{\bf b}}=\widehat{{\bf c}}$, for which $"\left[ \sigma _z\right] "
$ is given in Eq. (\ref{si64}), we can deduce the matrix forms of the
operators for the $x$ and $y$ components of spin by first obtaining the
ladder operators $\left[ \sigma _{+}\right] \;$and $\left[ \sigma
_{-}\right] .$ We obtain the ladder operators by their actions on the
vectors of $\left[ \sigma _z\right] .$ To make a distinction between vector
states and eigenvectors, we denote the latter by $[\xi _i]$, where $i$
stands for $+$, $0$ or $-$.$\;$We have

\begin{equation}
\left[ \sigma _{+}\right] [\xi _{+}]=0;\;\left[ \sigma _{+}\right] [\xi _0]=%
\sqrt{2}[\xi _{+}];\;\;\left[ \sigma _{+}\right] [\xi _{-}]=\sqrt{2}[\xi _0]
\label{si66}
\end{equation}
and

\begin{equation}
\left[ \sigma _{-}\right] [\xi _{-}]=0;\;\left[ \sigma _{-}\right] [\xi _0]=%
\sqrt{2}[\xi _{-}];\;\;\left[ \sigma _{-}\right] [\xi _{+}]=\sqrt{2}[\xi _0].
\label{si66a}
\end{equation}
We thus derive the standard formulas

\begin{equation}
\left[ \sigma _{+}\right] =\left( 
\begin{array}{ccc}
0 & \sqrt{2} & 0 \\ 
0 & 0 & \sqrt{2} \\ 
0 & 0 & 0
\end{array}
\right)  \label{si67}
\end{equation}
and

\begin{equation}
\left[ \sigma _{-}\right] =\left( 
\begin{array}{ccc}
0 & 0 & 0 \\ 
\sqrt{2} & 0 & 0 \\ 
0 & \sqrt{2} & 0
\end{array}
\right) .  \label{si68}
\end{equation}
From

\begin{equation}
\left[ \sigma _x\right] =\frac 12(\left[ \sigma _{+}\right] +\left[ \sigma
_{-}\right] )  \label{si69}
\end{equation}
and 
\begin{equation}
\lbrack \sigma _y]=-\frac i2(\left[ \sigma _{+}\right] -\left[ \sigma
_{-}\right] ),  \label{se70}
\end{equation}
we obtain

\begin{equation}
\left[ \sigma _x\right] =\frac 1{\sqrt{2}}\left( 
\begin{array}{ccc}
0 & 1 & 0 \\ 
1 & 0 & 1 \\ 
0 & 1 & 0
\end{array}
\right)  \label{se71}
\end{equation}
and 
\begin{equation}
\lbrack \sigma _y]=\frac 1{\sqrt{2}}\left( 
\begin{array}{ccc}
0 & -i & 0 \\ 
i & 0 & -i \\ 
0 & i & 0
\end{array}
\right) .  \label{se72}
\end{equation}

Thus, we have derived from first principles the operators for spin $1,$ just
as we derived the operators for spin $1/2$ from first principles in ref. $1$%
. Now, these spin matrices are clearly not the most general operators for
the description of spin measurements. To obtain these general forms, we need
the explicit expressions for the probability amplitudes so that we can
substitute them in the formulas Eq. (\ref{tw29}) - (\ref{th37}).

\subsection{\bf Explicit Expressions for the Probability Amplitudes}

The standard matrix operator for the spin is

\begin{equation}
\left[ {\bf \sigma }\right] =\widehat{{\bf i}}\left( 
\begin{array}{ccc}
1 & 0 & 0 \\ 
0 & 0 & 0 \\ 
0 & 0 & -1
\end{array}
\right) +\widehat{{\bf j}}\frac 1{\sqrt{2}}\left( 
\begin{array}{ccc}
0 & 1 & 0 \\ 
1 & 0 & 1 \\ 
0 & 1 & 0
\end{array}
\right) +\widehat{{\bf k}}\frac 1{\sqrt{2}}\left( 
\begin{array}{ccc}
0 & -i & 0 \\ 
i & 0 & -i \\ 
0 & i & 0
\end{array}
\right) .  \label{se73}
\end{equation}

We now take the dot product of this operator with the unit vector $\widehat{%
{\bf a}}$. If the polar angles of $\widehat{{\bf a}}$ are $(\theta ,\varphi
) $, then in Cartesian coordinates, $\widehat{{\bf a}}=(\sin \theta \cos
\varphi ,\sin \theta \sin \varphi ,\cos \theta ).$ A measurement of spin
along the new direction defined by the unit vector ${\bf a}$ gives the
values $0,\pm 1.$ We have

\begin{equation}
\lbrack {\bf \sigma }\cdot \widehat{{\bf a}}]=\left( 
\begin{array}{ccc}
\cos \theta & \frac 1{\sqrt{2}}\sin \theta e^{-i\varphi } & 0 \\ 
\frac 1{\sqrt{2}}\sin \theta e^{i\varphi } & 0 & \frac 1{\sqrt{2}}\sin
\theta e^{-i\varphi } \\ 
0 & \frac 1{\sqrt{2}}\sin \theta e^{i\varphi } & -\cos \theta
\end{array}
\right) .  \label{se74}
\end{equation}
The eigenvalues of this operator are $+1,\;0\;$and $-1$ with the respective
normalized eigenvectors

\begin{equation}
\lbrack \xi _{+}]=\left( 
\begin{array}{c}
\cos ^2\frac \theta 2e^{-i\varphi } \\ 
\frac 1{\sqrt{2}}\sin \theta \\ 
\sin ^2\frac \theta 2e^{i\varphi }
\end{array}
\right) ,  \label{se75}
\end{equation}
\begin{equation}
\lbrack \xi _0]=\left( 
\begin{array}{c}
-\frac 1{\sqrt{2}}\sin \theta e^{-i\varphi } \\ 
\cos \theta \\ 
\frac 1{\sqrt{2}}\sin \theta e^{i\varphi }
\end{array}
\right)  \label{se76}
\end{equation}
and

\begin{equation}
\lbrack \xi _{-}]=\left( 
\begin{array}{c}
-\sin ^2\frac \theta 2e^{-i\varphi } \\ 
\frac 1{\sqrt{2}}\sin \theta \\ 
-\cos ^2\frac \theta 2e^{i\varphi }
\end{array}
\right) .  \label{se77}
\end{equation}

The vectors Eqns. (\ref{se75}) - (\ref{se77}) are the eigenvectors resulting
from solving the eigenvalue equation Eq. (\ref{seven}). Therefore, they are
of the form $\chi (a_k;B_j),$ where $a_k$ is an eigenvalue of the matrix,
and corresponds to the state that precedes measurement, while $B_j$ is an
eigenvalue corresponding to another state. In general $B$ is not the
quantity being measured. The quantity $B$ is the spin projection along
another direction which is defined by some unit vector $\widehat{{\bf d}}$,
which for the moment is unidentified. We can see that in the vectors $[\xi
_{+}]$, $[\xi _0]$ and $[\xi _{-}],$ the elements are probability
amplitudes. Thus, for example, consider the elements of $[\xi _{+}].$ These
are probability amplitudes; they refer to measurements from the initial
state defined by the spin projection $+1$ with respect to the vector $%
\widehat{{\bf a}}$ to the states characterized by the final projections $+1$%
, $0$ and $-1$ with respect to the unknown unit vector $\widehat{{\bf d}}$.
Denoting the probability amplitudes by the symbol $\chi $, we deduce that

\begin{equation}
\chi ((+1)^{(\widehat{{\bf a}})};(+1)^{(\widehat{{\bf d}})})=\cos ^2\frac
\theta 2e^{-i\varphi },  \label{se78}
\end{equation}

\begin{equation}
\chi ((+1)^{(\widehat{{\bf a}})};0)^{(\widehat{{\bf d}})})=\frac 1{\sqrt{2}%
}\sin \theta  \label{se79}
\end{equation}
and 
\begin{equation}
\chi ((+1)^{(\widehat{{\bf a}})};(-1)^{(\widehat{{\bf d}})})=\sin ^2\frac
\theta 2e^{i\varphi }.  \label{ei80}
\end{equation}
Similarly, we deduce from $[\xi _0]$ and $[\xi _{-}]$ that 
\begin{equation}
\chi (0^{(\widehat{{\bf a}})};(+1)^{(\widehat{{\bf d}})})=-\frac 1{\sqrt{2}%
}\sin \theta e^{-i\varphi },  \label{ei81}
\end{equation}

\begin{equation}
\chi (0^{(\widehat{{\bf a}})};0)^{(\widehat{{\bf d}})})=\cos \theta ,
\label{ei82}
\end{equation}

\begin{equation}
\chi (0)^{(\widehat{{\bf a}})};(-1)^{(\widehat{{\bf d}})})=\frac 1{\sqrt{2}%
}\sin \theta e^{i\varphi },  \label{ei83}
\end{equation}
\begin{equation}
\chi ((-1)^{(\widehat{{\bf a}})};(+1)^{(\widehat{{\bf d}})})=-\sin ^2\frac
\theta 2e^{-i\varphi },  \label{ei84}
\end{equation}

\begin{equation}
\chi ((-1)^{(\widehat{{\bf a}})};0^{(\widehat{{\bf d}})})=\frac 1{\sqrt{2}%
}\sin \theta  \label{ei85}
\end{equation}
and 
\begin{equation}
\chi ((-1)^{(\widehat{{\bf a}})};(-1)^{(\widehat{{\bf d}})})=-\cos ^2\frac
\theta 2e^{i\varphi }  \label{ei86}
\end{equation}

These expressions are not necessarily the most general probability
amplitudes because $\widehat{{\bf d}}$ may not be an arbitrary vector. In
fact, since generalized probability amplitudes must contain angles
pertaining to two different directions, we can see that the probability
amplitudes Eqns. (\ref{se78}) - (\ref{ei86}) are not candidates for this
role. However, in view of the Land\'e formula Eq. (\ref{two}), we can use
Eqns. (\ref{se78}) - (\ref{ei86}) to eliminate the unknown direction $%
\widehat{{\bf d}}$ and hence deduce the general form of the probability
amplitudes. Thus, we introduce a new unit vector $\widehat{{\bf c}}$ defined
by the polar angles $(\theta ^{\prime },\varphi ^{\prime }).$ From the
eigenvectors of the operator $[{\bf \sigma }\cdot \widehat{{\bf c}}]$, we
deduce that

\begin{equation}
\chi ((+1)^{(\widehat{{\bf c}})};(+1)^{(\widehat{{\bf d}})})=\cos ^2\frac{%
\theta ^{\prime }}2e^{-i\varphi ^{\prime }},  \label{ei89}
\end{equation}

\begin{equation}
\chi ((+1)^{(\widehat{{\bf c}})};0^{(\widehat{{\bf d}})})=\frac 1{\sqrt{2}%
}\sin \theta ^{\prime },  \label{ni90}
\end{equation}

\begin{equation}
\chi ((+1)^{(\widehat{{\bf c}})};(-1)^{(\widehat{{\bf d}})})=\sin ^2\frac{%
\theta ^{\prime }}2e^{i\varphi ^{\prime }},  \label{ni91}
\end{equation}
\begin{equation}
\chi (0^{(\widehat{{\bf c}})};(+1)^{(\widehat{{\bf d}})})=-\frac 1{\sqrt{2}%
}\sin \theta ^{\prime }e^{-i\varphi ^{\prime }},  \label{ni92}
\end{equation}

\begin{equation}
\chi (0^{(\widehat{{\bf c}})};0^{(\widehat{{\bf d}})})=\cos \theta ^{\prime
},  \label{ni93}
\end{equation}

\begin{equation}
\chi (0^{(\widehat{{\bf c}})};(-1)^{(\widehat{{\bf d}})})=\frac 1{\sqrt{2}%
}\sin \theta ^{\prime }e^{i\varphi \prime },  \label{ni94}
\end{equation}
\begin{equation}
\chi ((-1)^{(\widehat{{\bf c}})};(+1)^{(\widehat{{\bf d}})})=-\sin ^2\frac{%
\theta ^{\prime }}2e^{-i\varphi ^{\prime }},  \label{ni95}
\end{equation}

\begin{equation}
\chi ((-1)^{(\widehat{{\bf c}})};0^{(\widehat{{\bf d}})})=\frac 1{\sqrt{2}%
}\sin \theta ^{\prime }  \label{ni96}
\end{equation}
and 
\begin{equation}
\chi ((-1)^{(\widehat{{\bf c}})};(-1)^{(\widehat{{\bf d}})})=-\cos ^2\frac{%
\theta ^{\prime }}2e^{i\varphi ^{\prime }}.  \label{ni97}
\end{equation}

We can now use the Land\'e expansion Eq. (\ref{two}) to find the general
probability amplitudes $\psi (m_i^{(\widehat{{\bf a}})};m_f^{(\widehat{{\bf c%
}})})$. Using the Hermiticity condition Eq. (\ref{one}) to reverse arguments
in the probability amplitudes Eqns. (\ref{ei89}) - (\ref{ni97}), and using
the fact that all three sets of probability amplitudes $\psi $, $\chi $ and $%
\phi $ have exactly the same form, we find that

\begin{eqnarray}
\psi ((+1)^{(\widehat{{\bf a}})};(+1)^{(\widehat{{\bf c}})}) &=&\chi ((+1)^{(%
\widehat{{\bf a}})};(+1)^{(\widehat{{\bf d}})})\phi ((+1)^{(\widehat{{\bf d}}%
)};(+1)^{(\widehat{{\bf c}})})+\chi ((+1)^{(\widehat{{\bf a}})};0^{(\widehat{%
{\bf d}})})\phi (0^{(\widehat{{\bf d}})};(+1)^{(\widehat{{\bf c}})}) 
\nonumber \\
&&+\chi ((+1)^{(\widehat{{\bf a}})};(-1)^{(\widehat{{\bf d}})})\phi ((-1)^{(%
\widehat{{\bf d}})};(+1)^{(\widehat{{\bf c}})})  \nonumber \\
&=&\cos ^2\frac \theta 2\cos ^2\frac{\theta ^{\prime }}2e^{-i(\varphi
-\varphi ^{\prime })}+\sin ^2\frac \theta 2\sin ^2\frac{\theta ^{\prime }}%
2e^{i(\varphi -\varphi ^{\prime })}+\frac 12\sin \theta \sin \theta ^{\prime
}.  \label{ni98}
\end{eqnarray}
The other probability amplitudes are

\begin{eqnarray}
\chi ((+1)^{(\widehat{{\bf a}})};0^{(\widehat{{\bf c}})}) &=&\frac 1{\sqrt{2}%
}[\sin ^2\frac \theta 2\sin \theta ^{\prime }e^{i(\varphi -\varphi ^{\prime
})}-\cos ^2\frac \theta 2\sin \theta ^{\prime }e^{-i(\varphi -\varphi
^{\prime })}  \nonumber \\
&&+\sin \theta \cos \theta ^{\prime }],  \label{ni99}
\end{eqnarray}

\begin{eqnarray}
\chi ((+1)^{(\widehat{{\bf a}})};(-1)^{(\widehat{{\bf c}})}) &=&\cos ^2\frac
\theta 2\sin ^2\frac{\theta ^{\prime }}2e^{-i(\varphi -\varphi ^{\prime
})}+\sin ^2\frac \theta 2\cos ^2\frac{\theta ^{\prime }}2e^{i(\varphi
-\varphi ^{\prime })}  \nonumber \\
&&-\frac 12\sin \theta \sin \theta ^{\prime },  \label{hu100}
\end{eqnarray}
\begin{eqnarray}
\chi (0^{(\widehat{{\bf a}})};(+1)^{(\widehat{{\bf c}})}) &=&\frac 1{\sqrt{2}%
}[-\sin \theta \cos ^2\frac{\theta ^{\prime }}2e^{-i(\varphi -\varphi
^{\prime })}+\sin \theta \sin ^2\frac{\theta ^{\prime }}2e^{i(\varphi
-\varphi ^{\prime })}  \nonumber \\
&&+\cos \theta \sin \theta ^{\prime }],  \label{hu101}
\end{eqnarray}
\begin{equation}
\chi (0^{(\widehat{{\bf a}})};0^{(\widehat{{\bf c}})})=\frac 12\sin \theta
\sin \theta ^{\prime }e^{-i(\varphi -\varphi ^{\prime })}+\frac 12\sin
\theta \sin \theta ^{\prime }e^{i(\varphi -\varphi ^{\prime })}+\cos \theta
\cos \theta ^{\prime },  \label{hu102}
\end{equation}
\begin{eqnarray}
\chi (0^{(\widehat{{\bf a}})};(-1)^{(\widehat{{\bf c}})}) &=&\frac 1{\sqrt{2}%
}[-\sin \theta \sin ^2\frac{\theta ^{\prime }}2e^{-i(\varphi -\varphi
^{\prime })}+\sin \theta \cos ^2\frac{\theta ^{\prime }}2e^{i(\varphi
-\varphi ^{\prime })}  \nonumber \\
&&-\cos \theta \sin \theta ^{\prime }],  \label{hu103}
\end{eqnarray}
\begin{eqnarray}
\chi ((-1)^{(\widehat{{\bf a}})};(+1)^{(\widehat{{\bf c}})}) &=&\sin ^2\frac
\theta 2\cos ^2\frac{\theta ^{\prime }}2e^{-i(\varphi -\varphi ^{\prime
})}+\cos ^2\frac \theta 2\sin ^2\frac{\theta ^{\prime }}2e^{i(\varphi
-\varphi ^{\prime })}  \nonumber \\
&&-\frac 12\sin \theta \sin \theta ^{\prime },  \label{hu104}
\end{eqnarray}
\begin{eqnarray}
\chi ((-1)^{(\widehat{{\bf a}})};0^{(\widehat{{\bf c}})}) &=&\frac 1{\sqrt{2}%
}[-\sin ^2\frac \theta 2\sin \theta ^{\prime }e^{-i(\varphi -\varphi
^{\prime })}+\cos ^2\frac \theta 2\sin \theta ^{\prime }e^{i(\varphi
-\varphi ^{\prime })}  \nonumber \\
&&-\sin \theta \cos \theta ^{\prime }]  \label{hu105}
\end{eqnarray}
and

\begin{eqnarray}
\chi ((-1)^{(\widehat{{\bf a}})};(-1)^{(\widehat{{\bf c}})}) &=&\sin ^2\frac
\theta 2\sin ^2\frac{\theta ^{\prime }}2e^{-i(\varphi -\varphi ^{\prime
})}+\cos ^2\frac \theta 2\cos ^2\frac{\theta ^{\prime }}2e^{i(\varphi
-\varphi ^{\prime })}  \nonumber \\
&&+\frac 12\sin \theta \sin \theta ^{\prime }.  \label{hu106}
\end{eqnarray}

Once having obtained the generalized expressions Eqns. (\ref{ni98}) - (\ref
{hu106}), we are able to deduce the vector $\widehat{{\bf d}}$ which we used
to obtain them. This vector is evidently $\widehat{{\bf k}}$, the unit
vector defining the $z$ direction. Thus if we replace $(\theta ,\varphi )$
by $(\theta =\varphi =0),$ in the generalized expressions, we recover Eqns. (%
\ref{se78}) - (\ref{ei86}). The only difference to note is that the
probability amplitudes Eqns. (\ref{ei84}), (\ref{ei85}) and Eq. (\ref{ei86})
corresponding to the initial state $-a$ are multiplied by the phase factor $%
-1$.

\subsection{\bf Verification of the General Expressions}

If the probability amplitudes Eqns. (\ref{ni98}) - (\ref{hu106}) are
correct, they should internally satisfy the Land\'e formula Eq. (\ref{two}).
Let $\widehat{{\bf e}}$ be a new unit vector defined by the polar angles $%
(\theta ^{\prime \prime },\varphi ^{\prime \prime }).$ Then according to
Eqns. (\ref{ni98}) - (\ref{hu106}) ,

\begin{eqnarray}
\chi ((+1)^{(\widehat{{\bf a}})};(+1)^{(\widehat{{\bf e}})}) &=&\cos ^2\frac
\theta 2\cos ^2\frac{\theta ^{\prime \prime }}2e^{-i(\varphi -\varphi
^{\prime \prime })}+\sin ^2\frac \theta 2\sin ^2\frac{\theta ^{\prime \prime
}}2e^{i(\varphi -\varphi ^{\prime \prime })}  \nonumber \\
&&+\frac 12\sin \theta \sin \theta ^{\prime \prime },  \label{hu107} \\
&&\   \nonumber  \label{ei87}
\end{eqnarray}
\begin{eqnarray}
\chi ((+1)^{(\widehat{{\bf a}})};0^{(\widehat{{\bf e}})}) &=&\frac 1{\sqrt{2}%
}[\sin ^2\frac \theta 2\sin \theta ^{\prime \prime }e^{i(\varphi -\varphi
^{\prime \prime })}-\cos ^2\frac \theta 2\sin \theta ^{\prime \prime
}e^{-i(\varphi -\varphi ^{\prime \prime })}  \nonumber \\
&&+\sin \theta \cos \theta ^{\prime \prime }],  \label{hu108}
\end{eqnarray}

\begin{eqnarray}
\chi ((+1)^{(\widehat{{\bf a}})};(-1)^{(\widehat{{\bf e}})}) &=&\cos ^2\frac
\theta 2\sin ^2\frac{\theta ^{\prime \prime }}2e^{-i(\varphi -\varphi
^{\prime \prime })}+\sin ^2\frac \theta 2\cos ^2\frac{\theta ^{\prime \prime
}}2e^{i(\varphi -\varphi ^{\prime \prime })}  \nonumber \\
&&-\frac 12\sin \theta \sin \theta ^{\prime \prime },  \label{hu109}
\end{eqnarray}
\begin{eqnarray}
\chi (0^{(\widehat{{\bf a}})};(+1)^{(\widehat{{\bf e}})}) &=&\frac 1{\sqrt{2}%
}[-\sin \theta \cos ^2\frac{\theta ^{\prime \prime }}2e^{-i(\varphi -\varphi
^{\prime \prime })}+\sin \theta \sin ^2\frac{\theta ^{\prime \prime }}%
2e^{i(\varphi -\varphi ^{\prime \prime })}  \nonumber \\
&&+\cos \theta \sin \theta ^{\prime \prime }],  \label{hu110}
\end{eqnarray}
\begin{eqnarray}
\chi (0^{(\widehat{{\bf a}})};0^{(\widehat{{\bf e}})}) &=&\frac 12\sin
\theta \sin \theta ^{\prime \prime }e^{-i(\varphi -\varphi ^{\prime \prime
})}+\frac 12\sin \theta \sin \theta ^{\prime \prime }e^{i(\varphi -\varphi
^{\prime \prime })}  \nonumber \\
&&+\cos \theta \cos \theta ^{\prime \prime },  \label{hu111}
\end{eqnarray}
\begin{eqnarray}
\chi (0^{(\widehat{{\bf a}})};(-1)^{(\widehat{{\bf e}})}) &=&\frac 1{\sqrt{2}%
}[-\sin \theta \sin ^2\frac{\theta ^{\prime \prime }}2e^{-i(\varphi -\varphi
^{\prime \prime })}+\sin \theta \cos ^2\frac{\theta ^{\prime \prime }}%
2e^{i(\varphi -\varphi ^{\prime \prime })}  \nonumber \\
&&-\cos \theta \sin \theta ^{\prime \prime }],  \label{hu112}
\end{eqnarray}
\begin{eqnarray}
\chi ((-1)^{(\widehat{{\bf a}})};(+1)^{(\widehat{{\bf e}})}) &=&\sin
^2\theta /2\cos ^2\frac{\theta ^{\prime \prime }}2e^{-i(\varphi -\varphi
^{\prime \prime })}+\cos ^2\frac \theta 2\sin ^2\frac{\theta ^{\prime \prime
}}2e^{i(\varphi -\varphi ^{\prime \prime })}  \nonumber \\
&&-\frac 12\sin \theta \sin \theta ^{\prime \prime },  \label{hu113}
\end{eqnarray}
\begin{eqnarray}
\chi ((-1)^{(\widehat{{\bf a}})};0^{(\widehat{{\bf e}})}) &=&\frac 1{\sqrt{2}%
}[-\sin ^2\frac \theta 2\sin \theta ^{\prime \prime }e^{i(\varphi -\varphi
^{\prime \prime })}+\cos ^2\frac \theta 2\sin \theta ^{\prime \prime
}e^{-i(\varphi -\varphi ^{\prime \prime })}  \nonumber \\
&&-\sin \theta \cos \theta ^{\prime \prime }  \label{hu114}
\end{eqnarray}
and

\begin{eqnarray}
\chi ((-1)^{(\widehat{{\bf a}})};(-1)^{(\widehat{{\bf e}})}) &=&\sin ^2\frac
\theta 2\sin ^2\frac{\theta ^{\prime \prime }}2e^{-i(\varphi -\varphi
^{\prime \prime })}+\cos ^2\frac \theta 2\cos ^2\frac{\theta ^{\prime \prime
}}2e^{i(\varphi -\varphi ^{\prime \prime })}  \nonumber \\
&&+\frac 12\sin \theta \sin \theta ^{\prime \prime }.  \label{hu115}
\end{eqnarray}

We should be able to obtain all the probability amplitudes for measurements
from $\widehat{{\bf a}}$ to $\widehat{{\bf c}}$ by using these expressions
in the Land\'e expansion. Making use of the Hermiticity condition Eqn. (\ref
{one}), we find that

\begin{eqnarray}
\chi ((+1)^{(\widehat{{\bf c}})};(+1)^{(\widehat{{\bf e}})}) &=&\chi ((+1)^{(%
\widehat{{\bf c}})};(+1)^{(\widehat{{\bf a}})})\chi ((+1)^{(\widehat{{\bf a}}%
)};(+1)^{(\widehat{{\bf e}})})+\chi ((+1)^{(\widehat{{\bf c}})};0^{(\widehat{%
{\bf a}})})\chi (0^{(\widehat{{\bf a}})};(+1)^{(\widehat{{\bf e}})}) 
\nonumber \\
&&+\chi ((+1)^{(\widehat{{\bf c}})};(-1)^{(\widehat{{\bf a}})})\chi ((-1)^{(%
\widehat{{\bf a}})};(+1)^{(\widehat{{\bf e}})})  \nonumber \\
\ &=&\cos ^2\frac{\theta ^{\prime }}2\cos ^2\frac{\theta ^{\prime \prime }}%
2e^{-i(\varphi ^{\prime }-\varphi ^{\prime \prime })}+\sin ^2\frac{\theta
^{\prime }}2\sin ^2\frac{\theta ^{\prime \prime }}2e^{i(\varphi ^{\prime
}-\varphi ^{\prime \prime })}  \nonumber \\
&&+\frac 12\sin \theta ^{\prime }\sin \theta ^{\prime \prime }  \label{hu116}
\end{eqnarray}
the same result which is found by directly by putting the appropriate angles
in the relevant formula Eq. (\ref{ni98}). Thus, the Land\'e expansion is
satisfied for this probability amplitude: in fact, it is satisfied for all
others in the set $\chi (m_i^{(\widehat{{\bf c}})};m_f^{(\widehat{{\bf e}}%
)}) $ . Therefore, the probability amplitudes are correct and consistent.

\subsection{\bf Expressions for the Probabilities}

The probabilities corresponding to the probability amplitudes Eqns. (\ref
{ni98}) - (\ref{hu106}) are

\begin{eqnarray}
P((+1)^{(\widehat{{\bf a}})};(+1)^{(\widehat{{\bf c}})}) &=&\left| \chi
((+1)^{(\widehat{{\bf a}})};(+1)^{(\widehat{{\bf c}})})\right| ^2=[\cos
^2\frac \theta 2\cos ^2\frac{\theta ^{\prime }}2+\sin ^2\frac \theta 2\sin ^2%
\frac{\theta ^{\prime }}2  \nonumber \\
&&+\frac 12\sin \theta \sin \theta ^{\prime }\cos (\varphi ^{\prime
}-\varphi )]^2,  \label{hu117}
\end{eqnarray}

\begin{equation}
P((+1);0^{(\widehat{{\bf c}})})=\frac 12[1-(\cos \theta \cos \theta ^{\prime
}+\sin \theta \sin \theta ^{\prime }\cos (\varphi ^{\prime }-\varphi ))^2],
\label{hu118}
\end{equation}
\begin{eqnarray}
P((+1)^{(\widehat{{\bf a}})};(-1)^{(\widehat{{\bf c}})}) &=&[\cos ^2\frac
\theta 2\sin ^2\frac{\theta ^{\prime }}2+\sin ^2\frac \theta 2\cos ^2\frac{%
\theta ^{\prime }}2  \nonumber \\
&&-\frac 12\sin \theta \sin \theta ^{\prime }\cos (\varphi ^{\prime
}-\varphi )]^2,  \label{hu119}
\end{eqnarray}
\begin{equation}
P(0^{(\widehat{{\bf a}})};0^{(\widehat{{\bf c}})})=[\cos \theta \cos \theta
^{\prime }+\sin \theta \sin \theta ^{\prime }\cos (\varphi ^{\prime
}-\varphi )]^2,  \label{hu120}
\end{equation}
\begin{equation}
P(0^{(\widehat{{\bf a}})};(+1)^{(\widehat{{\bf c}})})=P((+1)^{(\widehat{{\bf %
a}})};0^{(\widehat{{\bf c}})}),  \label{hu121}
\end{equation}
\begin{equation}
P(0^{(\widehat{{\bf a}})};(-1)^{(\widehat{{\bf c}})})=P((+1)^{(\widehat{{\bf %
a}})};0^{(\widehat{{\bf c}})}),  \label{hu122}
\end{equation}
\begin{equation}
P((-1)^{(\widehat{{\bf a}})};(+1)^{(\widehat{{\bf c}})})=P((+1)^{(\widehat{%
{\bf a}})};(-1)^{(\widehat{{\bf c}})}),  \label{hu123}
\end{equation}
\begin{equation}
P((-1)^{(\widehat{{\bf a}})};0^{(\widehat{{\bf c}})})=P((+1)^{(\widehat{{\bf %
a}})};0^{(\widehat{{\bf c}})})  \label{hu124}
\end{equation}
and 
\begin{equation}
P((-1)^{(\widehat{{\bf a}})};(-1)^{(\widehat{{\bf c}})})=P((+1)^{(\widehat{%
{\bf a}})};(+1)^{(\widehat{{\bf c}})}).  \label{hu125}
\end{equation}
The probabilities corresponding to a particular initial state should sum to
unity. We verify that

\begin{equation}
\sum_{n=1}^3P(m_i^{(\widehat{{\bf a}})};m_n^{(\widehat{{\bf c}}%
)})=1,\;\;(i=1,2,3).  \label{hu126}
\end{equation}

\subsection{{\bf Explicit Formulas For The }${\bf z}${\bf \ Component Of The
Spin Operator}}

We are now in a position to use the defining expressions Eqns. (\ref{tw29})
- (\ref{th37}), to deduce the expressions for the elements of the
generalized ''$z$ component'' of the spin operator. Since this operator is
referred to the direction $\widehat{{\bf c}}$, we now denote it by $[\sigma
_{\widehat{{\bf c}}}]$. In the defining expressions, we have probability
amplitudes of the form $\phi (m_i^{(\widehat{{\bf b}})};m_f^{(\widehat{{\bf c%
}})}).$ Let the angles of $\widehat{{\bf b}}$ be $(\theta ,\varphi )$ and
those of $\widehat{{\bf c}}$ be $(\theta ^{\prime },\varphi ^{\prime }).$ We
find that

\begin{equation}
(\sigma _{\widehat{{\bf c}}})_{11}=R_{11}=\cos \theta \cos \theta ^{\prime
}+\sin \theta \sin \theta ^{\prime }\cos (\varphi -\varphi ^{\prime }),
\label{hu129}
\end{equation}
\begin{eqnarray}
(\sigma _{\widehat{{\bf c}}})_{12} &=&R_{12}=\frac 1{\sqrt{2}}[-\sin \theta
\cos \theta ^{\prime }+\cos \theta \sin \theta ^{\prime }\cos (\varphi
-\varphi ^{\prime })  \nonumber \\
&&+i\sin \theta ^{\prime }\sin (\varphi -\varphi ^{\prime })],  \label{hu130}
\end{eqnarray}
\begin{equation}
(\sigma _{\widehat{{\bf c}}})_{13}=(\sigma _z)_{22}=(\sigma _z)_{31}=0,
\label{hu131}
\end{equation}
\begin{eqnarray}
(\sigma _{\widehat{{\bf c}}})_{21} &=&\frac 1{\sqrt{2}}[-\sin \theta \cos
\theta ^{\prime }+\cos \theta \sin \theta ^{\prime }\cos (\varphi -\varphi
^{\prime })  \nonumber \\
&&-i\sin \theta ^{\prime }\sin (\varphi -\varphi ^{\prime })],  \label{hu132}
\end{eqnarray}

\begin{equation}
(\sigma _{\widehat{{\bf c}}})_{23}=(\sigma _{\widehat{{\bf c}}})_{12},
\label{hu133}
\end{equation}

\begin{equation}
(\sigma _{\widehat{{\bf c}}})_{32}=(\sigma _{\widehat{{\bf c}}})_{21}
\label{hu134}
\end{equation}
and 
\begin{equation}
(\sigma _{\widehat{{\bf c}}})_{33}=-(\sigma _{\widehat{{\bf c}}})_{11}.
\label{hu135}
\end{equation}

When we set $\widehat{{\bf b}}=\widehat{{\bf k}}$ in the expressions above
we obtain the elements of the operator Eq. (\ref{se77}). The only difference
to note is that due to the change of notation in Subsection $3.7$, the
angles that have to be set to zero are $\theta $ and $\varphi .$

We now need the eigenvectors of the generalized operator. According to the
matrix eigenvalue equation Eq. (\ref{seven}), the elements of the
eigenvectors are probability amplitudes. Therefore, they have the forms of
the generalized probability amplitudes. When we examine Eqns. (\ref{th38}) -
(\ref{si63}), we discover that the case for which $\widehat{{\bf a}}=%
\widehat{{\bf c}}$ while $\widehat{{\bf b}}$ is arbitrary corresponds to the
situation in which the operator has its most general form, while at the same
time the vectors are eigenvectors of the operator. Hence the eigenvectors of
the generalized operator correspond to the case where the initial state in
each eigenvector belongs to the vector $\widehat{{\bf c}}$, while the final
states belong to the vector $\widehat{{\bf b}}$. This is why in order to
recover the special vectors Eqns. (\ref{se75}) - (\ref{se77}), we set $%
\widehat{{\bf b}}=\widehat{{\bf k}}$. We keep in mind that $\widehat{{\bf b}}
$ defines the intermediate observable.

The elements of the generalized eigenvectors are probability amplitudes of
the form $\chi (m_i^{(\widehat{{\bf c}})};m_f^{\widehat{{\bf b}}}).$ Since
they are referred to the arbitrary direction $\widehat{{\bf c}},$ the
eigenvectors are now denoted by $[\xi _{\widehat{{\bf c}}}^{(+)}],$ $[\xi _{%
\widehat{{\bf c}}}^{(0)}]$ and $[\xi _{\widehat{{\bf c}}}^{(-)}]$. They are

\begin{equation}
\lbrack \xi _{\widehat{{\bf c}}}^{(+)}]=\left( 
\begin{array}{c}
\chi ((+1)^{(\widehat{{\bf c}})};(+1)^{(\widehat{{\bf b}})}) \\ 
\chi ((+1)^{(\widehat{{\bf c}})};0^{(\widehat{{\bf b}})}) \\ 
\chi ((+1)^{(\widehat{{\bf c}})};(-1)^{(\widehat{{\bf b}})})
\end{array}
\right) =\left( 
\begin{array}{c}
\cos ^2\frac{\theta ^{\prime }}2\cos ^2\frac \theta 2e^{-i(\varphi ^{\prime
}-\varphi )}+\sin ^2\frac{\theta ^{\prime }}2\sin ^2\frac \theta
2e^{i(\varphi ^{\prime }-\varphi )}+\frac 12\sin \theta ^{\prime }\sin \theta
\\ 
\frac 1{\sqrt{2}}[-\cos ^2\frac{\theta ^{\prime }}2\sin \theta e^{-i(\varphi
^{\prime }-\varphi )}+\sin ^2\frac{\theta ^{\prime }}2\sin \theta
e^{i(\varphi ^{\prime }-\varphi )}+\sin \theta ^{\prime }\cos \theta ] \\ 
\cos ^2\frac{\theta ^{\prime }}2\sin ^2\frac \theta 2e^{-i(\varphi ^{\prime
}-\varphi )}+\sin ^2\frac{\theta ^{\prime }}2\cos ^2\frac \theta
2e^{i(\varphi ^{\prime }-\varphi )}-\frac 12\sin \theta ^{\prime }\sin \theta
\end{array}
\right) ,  \label{hu136}
\end{equation}

\begin{equation}
\lbrack \xi _{\widehat{{\bf c}}}^{(0)}]=\left( 
\begin{array}{c}
\chi (0^{(\widehat{{\bf c}})};(+1)^{(\widehat{{\bf b}})}) \\ 
\chi (0^{(\widehat{{\bf c}})};0^{(\widehat{{\bf b}})}) \\ 
\chi (0^{(\widehat{{\bf c}})};(-1)^{(\widehat{{\bf b}})})
\end{array}
\right) =\left( 
\begin{array}{c}
\frac 1{\sqrt{2}}[-\cos ^2\frac \theta 2\sin \theta ^{\prime }e^{-i(\varphi
^{\prime }-\varphi )}+\sin ^2\frac \theta 2\sin \theta ^{\prime
}e^{i(\varphi ^{\prime }-\varphi )}+\sin \theta \cos \theta ^{\prime }] \\ 
\frac 12\sin \theta \sin \theta ^{\prime }e^{-i(\varphi ^{\prime }-\varphi
)}+\frac 12\sin \theta \sin \theta ^{\prime }e^{i(\varphi ^{\prime }-\varphi
)}+\cos \theta \cos \theta ^{\prime } \\ 
\frac 1{\sqrt{2}}[-\sin ^2\frac \theta 2\sin \theta ^{\prime }e^{-i(\varphi
^{\prime }-\varphi )}+\cos ^2\frac \theta 2\sin \theta ^{\prime
}e^{i(\varphi ^{\prime }-\varphi )}-\sin \theta \cos \theta ^{\prime }]
\end{array}
\right)  \label{hu137}
\end{equation}
and 
\begin{equation}
\lbrack \xi _{\widehat{{\bf c}}}^{(-)}]=\left( 
\begin{array}{c}
\chi ((-1)^{(\widehat{{\bf c}})};(+1)^{(\widehat{{\bf b}})}) \\ 
\chi ((-1)^{(\widehat{{\bf c}})};0^{(\widehat{{\bf b}})}) \\ 
\chi ((-1)^{(\widehat{{\bf c}})};(-1)^{(\widehat{{\bf b}})})
\end{array}
\right) =\left( 
\begin{array}{c}
\sin ^2\frac{\theta ^{\prime }}2\cos ^2\frac \theta 2e^{-i(\varphi ^{\prime
}-\varphi )}+\cos ^2\frac{\theta ^{\prime }}2\sin ^2\frac \theta
2e^{i(\varphi ^{\prime }-\varphi )}-\frac 12\sin \theta ^{\prime }\sin \theta
\\ 
\frac 1{\sqrt{2}}[-\sin ^2\frac{\theta ^{\prime }}2\sin \theta e^{-i(\varphi
^{\prime }-\varphi )}+\cos ^2\frac{\theta ^{\prime }}2\sin \theta
e^{i(\varphi ^{\prime }-\varphi )}-\sin \theta ^{\prime }\cos \theta ] \\ 
\sin ^2\frac{\theta ^{\prime }}2\sin ^2\frac \theta 2e^{-i(\varphi ^{\prime
}-\varphi )}+\cos ^2\frac{\theta ^{\prime }}2\cos ^2\frac \theta
2e^{i(\varphi ^{\prime }-\varphi )}+\frac 12\sin \theta ^{\prime }\sin \theta
\end{array}
\right) .  \label{hu138}
\end{equation}
Direct substitution verifies that these vectors do satisfy the eigenvalue
equations

\begin{equation}
\left[ \sigma _z\right] [\xi _{\widehat{{\bf c}}}^{(+)}]=(+1)[\xi _{\widehat{%
{\bf c}}}^{(+)}],  \label{hu139}
\end{equation}

\begin{equation}
\left[ \sigma _z\right] [\xi _{\widehat{{\bf c}}}^{(0)}]=0  \label{hu140}
\end{equation}
and 
\begin{equation}
\left[ \sigma _z\right] [\xi _{\widehat{{\bf c}}}^{(-)}]=(-1)[\xi _{\widehat{%
{\bf c}}}^{(-)}].  \label{hu141}
\end{equation}

\subsection{{\bf Generalized Expressions for the }$x${\bf \ and }$y${\bf \
Components of the Spin}}

We now seek the operators $[\sigma _x]$ and $[\sigma _y].$ The most obvious
way to obtain them is to use the ladder operators, which we can derive by
their actions on the eigenvectors of $[\sigma _{\widehat{{\bf c}}}]$. This
is the procedure followed in ref. $1$ for spin $1/2$. However, it proved
time-consuming in that case, and would be even more so in this case. Another
method would be to used rotations. However, neither of these two methods is
necessary. Fortunately, as we have shown in ref. $2$, we can get the
operators $[\sigma _x]$ and $[\sigma _y]$ simply by changing arguments in
the expression for the generalized operator $[\sigma _{\widehat{{\bf c}}}].$
By the same change of arguments, the vectors of $[\sigma _{\widehat{{\bf c}}%
}]$ transform to the vectors of $[\sigma _x]$ or $[\sigma _y].$ We proved
this procedure by applying it to the case of spin $1/2$ for which the
operators $[\sigma _x]$ and $[\sigma _y]$ were already known [$2$].

For the case of spin $1/2$, we found that the transformation $\theta
^{\prime }\rightarrow \theta ^{\prime }-\pi /2$ applied to $[\sigma _{%
\widehat{{\bf c}}}]$ and its eigenvectors yielded $[\sigma _x]$ and its
eigenvectors. The transformations $\theta ^{\prime }=\pi /2$ and $\varphi
^{\prime }\rightarrow \varphi ^{\prime }-\pi /2$ gave $[\sigma _y]$ and its
eigenvectors. When we try these transformations in this case, we find that
we obtain the wrong results for both $[\sigma _x]$ and $[\sigma _y].$ We
know this because if we set $\theta =\theta ^{\prime }$ and $\varphi
=\varphi ^{\prime }$ in these operators in order to recover the standard
results Eqns. (\ref{se71}) and (\ref{se72}), we fail to do so. However a
little trial and error shows that if we set $\theta ^{\prime }\rightarrow
\theta ^{\prime }+\pi /2,$we do obtain $[\sigma _x]$ and its eigenvectors.
Similarly, $\theta ^{\prime }=\pi /2$ and $\varphi ^{\prime }\rightarrow
\varphi ^{\prime }+\pi /2$ lead to the correct results for $[\sigma _y]$ and
its eigenvectors. The standard results are obtained in the limit $\theta
=\theta ^{\prime }$ and $\varphi =\varphi ^{\prime }$.

Using these transformations, the elements of $[\sigma _x]$ are found to be

\begin{equation}
(\sigma _x)_{11}=-\cos \theta \sin \theta ^{\prime }+\sin \theta \cos \theta
^{\prime }\cos (\varphi -\varphi ^{\prime }),  \label{hu142}
\end{equation}

\begin{eqnarray}
(\sigma _x)_{12} &=&\frac 1{\sqrt{2}}[\sin \theta \sin \theta ^{\prime
}+\cos \theta \cos \theta ^{\prime }\cos (\varphi -\varphi ^{\prime }) 
\nonumber \\
&&+i\cos \theta ^{\prime }\sin (\varphi -\varphi ^{\prime })],  \label{hu143}
\end{eqnarray}

\begin{eqnarray}
(\sigma _x)_{21} &=&\frac 1{\sqrt{2}}[\sin \theta \sin \theta ^{\prime
}+\cos \theta \cos \theta ^{\prime }\cos (\varphi -\varphi ^{\prime }) 
\nonumber \\
&&-i\cos \theta ^{\prime }\sin (\varphi -\varphi ^{\prime })],  \label{hu144}
\\
(\sigma _x)_{23} &=&(\sigma _x)_{12},  \label{hu145} \\
(\sigma _x)_{32} &=&(\sigma _x)_{21},  \label{hu146} \\
(\sigma _x)_{33} &=&-(\sigma _x)_{11}  \label{hu147}
\end{eqnarray}
and 
\begin{equation}
(\sigma _x)_{13}=(\sigma _x)_{22}=(\sigma _x)_{31}=0.  \label{hu148}
\end{equation}

The eigenvector corresponding to the eigenvalue $+1$ is

\begin{equation}
\lbrack \xi _x^{(+)}]=\left( 
\begin{array}{c}
\frac 12(1-\sin \theta ^{\prime })\cos ^2\frac \theta 2e^{-i(\varphi
^{\prime }-\varphi )}+\frac 12(1+\sin \theta ^{\prime })\sin ^2\frac \theta
2e^{i(\varphi ^{\prime }-\varphi )}+\frac 12\sin \theta \cos \theta ^{\prime
} \\ 
\frac 1{\sqrt{2}}[-\frac 12(1-\sin \theta ^{\prime })\sin \theta
e^{-i(\varphi ^{\prime }-\varphi )}+\frac 12(1+\sin \theta ^{\prime })\sin
\theta e^{i(\varphi ^{\prime }-\varphi )}+\cos \theta \cos \theta ^{\prime }]
\\ 
\frac 12(1-\sin \theta ^{\prime })\sin ^2\frac \theta 2e^{-i(\varphi
^{\prime }-\varphi )}+\frac 12(1+\sin \theta ^{\prime })\cos ^2\frac \theta
2e^{i(\varphi ^{\prime }-\varphi )}-\frac 12\sin \theta \cos \theta ^{\prime
}
\end{array}
\right) .  \label{hu149}
\end{equation}

The eigenvector corresponding to the eigenvalue $0$ is

\begin{equation}
\lbrack \xi _x^{(0)}]=\left( 
\begin{array}{c}
\frac 1{\sqrt{2}}[-\cos \theta ^{\prime }\cos ^2\frac \theta 2e^{-i(\varphi
^{\prime }-\varphi )}+\cos \theta ^{\prime }\sin ^2\frac \theta
2e^{i(\varphi ^{\prime }-\varphi )}-\sin \theta \sin \theta ^{\prime }] \\ 
\frac 12\cos \theta ^{\prime }\sin \theta e^{-i(\varphi ^{\prime }-\varphi
)}+\frac 12\cos \theta ^{\prime }\sin \theta e^{i(\varphi ^{\prime }-\varphi
)}-\cos \theta \sin \theta ^{\prime } \\ 
\frac 1{\sqrt{2}}[-\cos \theta ^{\prime }\sin ^2\frac \theta 2e^{-i(\varphi
^{\prime }-\varphi )}+\cos \theta ^{\prime }\cos ^2\frac \theta
2e^{i(\varphi ^{\prime }-\varphi )}+\sin \theta \sin \theta ^{\prime }]
\end{array}
\right) .  \label{hu150}
\end{equation}

The eigenvector corresponding to the eigenvalue $-1$ is

\begin{equation}
\lbrack \xi _x^{(-)}]=\left( 
\begin{array}{c}
\frac 12(1+\sin \theta ^{\prime })\cos ^2\frac \theta 2e^{-i(\varphi
^{\prime }-\varphi )}+\frac 12(1-\sin \theta ^{\prime })\sin ^2\frac \theta
2e^{i(\varphi ^{\prime }-\varphi )}-\frac 12\sin \theta \cos \theta ^{\prime
} \\ 
\frac 1{\sqrt{2}}[-\frac 12(1+\sin \theta ^{\prime })\sin \theta
e^{-i(\varphi ^{\prime }-\varphi )}+\frac 12(1-\sin \theta ^{\prime })\sin
\theta e^{i(\varphi ^{\prime }-\varphi )}-\cos \theta \cos \theta ^{\prime }]
\\ 
\frac 12(1+\sin \theta ^{\prime })\sin ^2\frac \theta 2e^{-i(\varphi
^{\prime }-\varphi )}+\frac 12(1-\sin \theta ^{\prime })\cos ^2\frac \theta
2e^{i(\varphi ^{\prime }-\varphi )}+\frac 12\sin \theta \cos \theta ^{\prime
}
\end{array}
\right) .  \label{hu151}
\end{equation}

We confirm that these eigenvectors satisfy the eigenvalue equations

\begin{equation}
\lbrack \sigma _x][\xi _x^{(+)}]=(+1)[\xi _x^{(+)}],  \label{hu152}
\end{equation}

\begin{equation}
\lbrack \sigma _x][\xi _x^{(0)}]=0  \label{hu153}
\end{equation}
and

\begin{equation}
\lbrack \sigma _x][\xi _x^{(-)}]=(-1)[\xi _x^{(-)}].  \label{hu154}
\end{equation}

The elements of $[\sigma _y]$ are found to be

\begin{eqnarray}
(\sigma _y)_{11} &=&\sin \theta \sin (\varphi -\varphi ^{\prime }),
\label{hu155} \\
(\sigma _y)_{12} &=&\frac 1{\sqrt{2}}[\cos \theta \sin (\varphi -\varphi
^{\prime })-i\cos (\varphi -\varphi ^{\prime })],  \label{hu156} \\
(\sigma _y)_{13} &=&(\sigma _y)_{22}=(\sigma _y)_{31}=0,  \label{hu157} \\
(\sigma _y)_{21} &=&\frac 1{\sqrt{2}}[\cos \theta \sin (\varphi -\varphi
^{\prime })+i\cos (\varphi -\varphi ^{\prime })],  \label{hu158} \\
(\sigma _y)_{23} &=&(\sigma _y)_{12},  \label{hu159} \\
(\sigma _y)_{32} &=&(\sigma _y)_{21}  \label{hu160}
\end{eqnarray}
and 
\begin{equation}
(\sigma _y)_{33}=-(\sigma _y)_{11}.  \label{hu161}
\end{equation}
The eigenvector corresponding to the eigenvalue $+1$ is

\begin{equation}
\lbrack \xi _y^{(+)}]=\left( 
\begin{array}{c}
\frac 12[-i\cos ^2\frac \theta 2e^{-i(\varphi ^{\prime }-\varphi )}+i\sin
^2\frac \theta 2e^{i(\varphi ^{\prime }-\varphi )}+\sin \theta ] \\ 
\frac 1{\sqrt{2}}[\frac i2\sin \theta e^{-i(\varphi ^{\prime }-\varphi
)}+\frac i2\sin \theta e^{i(\varphi ^{\prime }-\varphi )}+\cos \theta ] \\ 
\frac 12[-i\sin ^2\frac \theta 2e^{-i(\varphi ^{\prime }-\varphi )}+i\cos
^2\frac \theta 2e^{i(\varphi ^{\prime }-\varphi )}-\sin \theta ]
\end{array}
\right) .  \label{hu162}
\end{equation}
The eigenvector corresponding to the eigenvalue $0$ is 
\begin{equation}
\lbrack \xi _y^{(0)}]=\left( 
\begin{array}{c}
\frac 1{\sqrt{2}}[i\cos ^2\frac \theta 2e^{-i(\varphi ^{\prime }-\varphi
)}+i\sin ^2\frac \theta 2e^{i(\varphi ^{\prime }-\varphi )}] \\ 
\frac 12[-i\sin \theta e^{-i(\varphi ^{\prime }-\varphi )}+i\sin \theta
e^{i(\varphi ^{\prime }-\varphi )}] \\ 
\frac 1{\sqrt{2}}[i\sin ^2\frac \theta 2e^{-i(\varphi ^{\prime }-\varphi
)}+i\cos ^2\frac \theta 2e^{i(\varphi ^{\prime }-\varphi )}]
\end{array}
\right) .  \label{hu163}
\end{equation}
The eigenvector corresponding to the eigenvalue $-1$ is

\begin{equation}
\lbrack \xi _y^{(-)}]=\left( 
\begin{array}{c}
\frac 12[-i\cos ^2\frac \theta 2e^{-i(\varphi ^{\prime }-\varphi )}+i\sin
^2\frac \theta 2e^{i(\varphi ^{\prime }-\varphi )}-\sin \theta ] \\ 
\frac 1{\sqrt{2}}[\frac i2\sin \theta e^{-i(\varphi ^{\prime }-\varphi
)}+\frac i2\sin \theta e^{i(\varphi ^{\prime }-\varphi )}-\cos \theta ] \\ 
\frac 12[-i\sin ^2\frac \theta 2e^{-i(\varphi ^{\prime }-\varphi )}+i\cos
^2\frac \theta 2e^{i(\varphi ^{\prime }-\varphi )}+\sin \theta ]
\end{array}
\right) .  \label{hu164}
\end{equation}

Direct substitution verifies that these eigenvectors satisfy the equations

\begin{equation}
\lbrack \sigma _y][\xi _y^{(+)}]=(+1)[\xi _y^{(+)}],  \label{hu165}
\end{equation}

\begin{equation}
\lbrack \sigma _y][\xi _y^{(0)}]=0  \label{hu166}
\end{equation}
and 
\begin{equation}
\lbrack \sigma _y][\xi _y^{(-)}]=(-1)[\xi _y^{(-)}].  \label{hu167}
\end{equation}

Calculation shows that the generalized operators $[\sigma _x]$, $[\sigma _y]$
and $[\sigma _z]$ satisfy the commutation relations

\begin{equation}
\lbrack [\sigma _x],[\sigma _{\widehat{{\bf c}}}]]=i[\sigma _z];\;\;[[\sigma
_y],[\sigma _{\widehat{{\bf c}}}]]=i[\sigma _x]\;\text{and\ [}[\sigma _{%
\widehat{{\bf c}}}],[\sigma _x]]=i[\sigma _y]  \label{hu168}
\end{equation}
and that the sum of their squares gives the matrix for the square of the
total spin:

\begin{equation}
\lbrack \sigma _x]^2+[\sigma _y]^2+[\sigma _{\widehat{{\bf c}}}]^2=\left( 
\begin{array}{ccc}
2 & 0 & 0 \\ 
0 & 2 & 0 \\ 
0 & 0 & 2
\end{array}
\right) .  \label{hu169}
\end{equation}

\subsection{\bf The Ladder Operators}

The elements of the raising operator $[\sigma _{+}]$ are

\begin{eqnarray}
(\sigma _{+})_{11} &=&-\cos \theta \sin \theta ^{\prime }+\sin \theta \cos
\theta ^{\prime }\cos (\varphi -\varphi ^{\prime })+i\sin \theta \sin
(\varphi -\varphi ^{\prime })  \label{hu170} \\
(\sigma _{+})_{12} &=&\frac 1{\sqrt{2}}[\sin \theta \sin \theta ^{\prime
}+\cos \theta \cos \theta ^{\prime }\cos (\varphi -\varphi ^{\prime })+\cos
(\varphi -\varphi ^{\prime })+i\cos \theta ^{\prime }\sin (\varphi -\varphi
^{\prime })  \nonumber \\
&&+i\cos \theta \sin (\varphi -\varphi ^{\prime })]  \label{hu171}
\end{eqnarray}

\begin{eqnarray}
(\sigma _{+})_{21} &=&\frac 1{\sqrt{2}}[\sin \theta \sin \theta ^{\prime
}+\cos \theta \cos \theta ^{\prime }\cos (\varphi -\varphi ^{\prime })-\cos
(\varphi -\varphi ^{\prime })-i\cos \theta ^{\prime }\sin (\varphi -\varphi
^{\prime })  \nonumber \\
&&+i\cos \theta \sin (\varphi -\varphi ^{\prime })]  \label{hu172}
\end{eqnarray}
\begin{equation}
(\sigma _{+})_{23}=(\sigma _{+})_{12}  \label{hu173}
\end{equation}
\begin{equation}
(\sigma _{+})_{32}=(\sigma _{+})_{21}  \label{hu174}
\end{equation}
\begin{equation}
(\sigma _{+})_{33}=\cos \theta \sin \theta ^{\prime }-\sin \theta \cos
\theta ^{\prime }\cos (\varphi -\varphi ^{\prime })-i\sin \theta \sin
(\varphi -\varphi ^{\prime })  \label{hu175}
\end{equation}
and 
\begin{equation}
(\sigma _{+})_{13}=(\sigma _{+})_{22}=(\sigma _{+})_{31}=0.  \label{hu176}
\end{equation}
The elements of the lowering operator $[\sigma _{-}]$ are

\begin{eqnarray}
(\sigma _{-})_{11} &=&-\cos \theta \sin \theta ^{\prime }+\sin \theta \cos
\theta ^{\prime }\cos (\varphi -\varphi ^{\prime })-i\sin \theta \sin
(\varphi -\varphi ^{\prime })  \label{hu177} \\
(\sigma _{-})_{12} &=&\frac 1{\sqrt{2}}[\sin \theta \sin \theta ^{\prime
}+\cos \theta \cos \theta ^{\prime }\cos (\varphi -\varphi ^{\prime })-\cos
(\varphi -\varphi ^{\prime })+i\cos \theta ^{\prime }\sin (\varphi -\varphi
^{\prime })  \nonumber \\
&&\ -i\cos \theta \sin (\varphi -\varphi ^{\prime })]  \label{hu178}
\end{eqnarray}

\begin{eqnarray}
(\sigma _{-})_{21} &=&\frac 1{\sqrt{2}}[\sin \theta \sin \theta ^{\prime
}+\cos \theta \cos \theta ^{\prime }\cos (\varphi -\varphi ^{\prime })+\cos
(\varphi -\varphi ^{\prime })-i\cos \theta ^{\prime }\sin (\varphi -\varphi
^{\prime })  \nonumber \\
&&\ -i\cos \theta \sin (\varphi -\varphi ^{\prime })]  \label{hu179}
\end{eqnarray}
\begin{equation}
(\sigma _{-})_{23}=(\sigma _{-})_{12}  \label{hu180}
\end{equation}
\begin{equation}
(\sigma _{-})_{32}=(\sigma _{-})_{21}  \label{hu181}
\end{equation}
\begin{equation}
(\sigma _{-})_{33}=\cos \theta \sin \theta ^{\prime }-\sin \theta \cos
\theta ^{\prime }\cos (\varphi -\varphi ^{\prime })+i\sin \theta \sin
(\varphi -\varphi ^{\prime })  \label{hu182}
\end{equation}
and 
\begin{equation}
(\sigma _{-})_{13}=(\sigma _{-})_{22}=(\sigma _{-})_{31}=0.  \label{hu183}
\end{equation}

We confirm that when these ladder operators act on the generalized vectors
of $[\sigma _{\widehat{{\bf c}}}],$ Eqns. \ref{hu109} - (\ref{hu111}), they
give

\begin{equation}
\lbrack \sigma _{+}][\xi _{\widehat{{\bf c}}}^{(+)}]=0;\;[\sigma _{+}][\xi _{%
\widehat{{\bf c}}}^{(0)}]=\sqrt{2}[\xi _{\widehat{{\bf c}}}^{(+)}%
];\;\;\;[\sigma _{+}][\xi _{\widehat{{\bf c}}}^{(-)}]=\sqrt{2}[\xi _{%
\widehat{{\bf c}}}^{(0)}]\;  \label{hu184}
\end{equation}
and

\begin{equation}
\lbrack \sigma _{-}][\xi _{\widehat{{\bf c}}}^{(-)}]=0;\;[\sigma _{-}][\xi _{%
\widehat{{\bf c}}}^{(0)}]=\sqrt{2}[\xi _{\widehat{{\bf c}}}^{(-)}%
];\;\;\;[\sigma _{-}][\xi _{\widehat{{\bf c}}}^{(+)}]=\sqrt{2}[\xi _{%
\widehat{{\bf c}}}^{(0)}].\;  \label{hu185}
\end{equation}

\subsection{\bf Generalized Form for the Square of the Spin}

We have seen that we do indeed obtain the elements of the matrix for the
square of the spin by the laborious procedure of squaring its components and
adding. Actually, this labour is unnecessary. We can directly infer that the
generalized form of this operator is diagonal.

The elements of the matrix for the square of the spin are given by the
general expressions Eqns. (\ref{tw29}) - (\ref{th37}). Now, the square of
the spin has the same value no matter what the spin projection is. For such
a quantity, the matrix operator is diagonal. We shall now prove that the
off-diagonal elements all vanish by showing this for $R_{12}.$

According to Eqn. (\ref{th30}),

\begin{eqnarray}
R_{12} &=&\phi ^{*}((+1)^{(\widehat{{\bf b}})};(+1)^{(\widehat{{\bf c}}%
)})\phi (0^{(\widehat{{\bf b}})};(+1)^{(\widehat{{\bf c}})})r_1+\phi
^{*}((+1)^{(\widehat{{\bf b}})};,0^{(\widehat{{\bf c}})})\phi (0^{(\widehat{%
{\bf b}})};0^{(\widehat{{\bf c}})})r_2  \nonumber \\
&&\ +\phi ^{*}((+1)^{(\widehat{{\bf b}})};(-1)^{(\widehat{{\bf c}})})\phi
(0^{(\widehat{{\bf b}})};(-1)^{(\widehat{{\bf c}})})r_3  \label{hu186}
\end{eqnarray}
Using the Hermiticity condition Eqn. (\ref{one}), and the fact that $%
r_1=r_2=r_3=r$, we see that

\begin{eqnarray}
R_{12} &=&[\phi (0^{(\widehat{{\bf b}})};(+1)^{(\widehat{{\bf c}})})\phi
((+1)^{(\widehat{{\bf c}})};(+1)^{(\widehat{{\bf b}})})+\phi (0^{(\widehat{%
{\bf b}})};0^{(\widehat{{\bf c}})})\phi (0^{(\widehat{{\bf c}})};(+1)^{(%
\widehat{{\bf b}})}))  \nonumber \\
&&+\phi (0^{(\widehat{{\bf b}})};(-1)^{(\widehat{{\bf c}})})\phi ((-1)^{(%
\widehat{{\bf c}})};(+1)^{(\widehat{{\bf b}})}]r.  \label{hu187}
\end{eqnarray}
Thus, according to the Land\'e expansion Eqn. (\ref{two}), Eqn. (\ref{hu187}%
) is

\begin{equation}
R_{12}=\phi (0^{(\widehat{{\bf b}})};(+1)^{(\widehat{{\bf b}})})r=0.
\label{hu188}
\end{equation}

The same reasoning shows that all the other off-diagonal elements vanish. On
the other hand, if we neglect the eigenvalue which multiplies each one, the
diagonal elements are just the sums of all the probabilities corresponding
to the same initial state. Thus for example,

\begin{equation}
R_{11}=[\left| \phi ((+1)^{(\widehat{{\bf b}})};(+1)^{(\widehat{{\bf c}}%
)})\right| ^2+\left| \phi ((+1)^{(\widehat{{\bf b}})};0^{(\widehat{{\bf c}}%
)})\right| ^2+\left| \phi ((+1)^{(\widehat{{\bf b}})};(-1)^{(\widehat{{\bf c}%
})})\right| ^2]r=r.  \label{hu189}
\end{equation}
Each diagonal element therefore equals the eigenvalue. In consequence, we
find that

\begin{equation}
\lbrack \sigma ^2]=\left( 
\begin{array}{ccc}
2 & 0 & 0 \\ 
0 & 2 & 0 \\ 
0 & 0 & 2
\end{array}
\right)  \label{hu500}
\end{equation}
a result we obtained earlier by squaring the components of the spin and
adding.

\subsection{\bf Further properties of the probability amplitudes}

Apart from the Hermiticity condition, the probability amplitudes satisfy the
following relations. Given two reference directions defined by the unit
vectors $\widehat{{\bf a}}$ and $\widehat{{\bf c}}$, we readily verify that 
\begin{equation}
\chi ^{*}((+1)^{(\widehat{{\bf a}})};(+1)^{(\widehat{{\bf c}})})=\chi
((-1)^{(\widehat{{\bf a}})};(-1)^{(\widehat{{\bf c}})}),  \label{hu190}
\end{equation}

\begin{eqnarray}
\chi ^{*}((+1)^{(\widehat{{\bf a}})};0^{(\widehat{{\bf c}})}) &=&-\chi
((-1);,0^{(\widehat{{\bf c}})}),  \label{hu191} \\
\chi ^{*}((+1)^{(\widehat{{\bf a}})};(-1)^{(\widehat{{\bf c}})}) &=&\chi
((-1)^{(\widehat{{\bf a}})};,(+1)^{(\widehat{{\bf c}})}),  \label{hu192} \\
\chi ^{*}(0^{(\widehat{{\bf a}})};(+1)^{(\widehat{{\bf c}})}) &=&-\chi (0^{(%
\widehat{{\bf a}})};,(-1)^{(\widehat{{\bf c}})}),  \label{hu193} \\
\chi ^{*}(0^{(\widehat{{\bf a}})};0^{(\widehat{{\bf c}})}) &=&\chi (0^{(%
\widehat{{\bf a}})};0^{(\widehat{{\bf c}})}),  \label{hu194} \\
\chi ^{*}(0^{(\widehat{{\bf a}})};(-1)^{(\widehat{{\bf c}})}) &=&-\chi (0^{(%
\widehat{{\bf a}})};(+1)^{(\widehat{{\bf c}})}),  \label{hu195} \\
\chi ^{*}((-1)^{(\widehat{{\bf a}})};(+1)^{(\widehat{{\bf c}})}) &=&\chi
((+1)^{(\widehat{{\bf a}})};(-1)^{(\widehat{{\bf c}})}),  \label{hu196} \\
\chi ^{*}((-1)^{(\widehat{{\bf a}})};0)^{(\widehat{{\bf c}})}) &=&-\chi
((+1)^{(\widehat{{\bf a}})};0^{(\widehat{{\bf c}})})  \label{hu197}
\end{eqnarray}
and 
\begin{equation}
\chi ^{*}((-1)^{(\widehat{{\bf a}})};(-1)^{(\widehat{{\bf c}})})=\chi
((+1)^{(\widehat{{\bf a}})};(+1)^{(\widehat{{\bf c}})}).  \label{hu198}
\end{equation}

\section{{\bf Results for Arbitrary Values of }$J$}

The method we have presented of obtaining generalized spin vectors and
operators has now been applied to the cases of spin $1/2$ and spin $1$. It
is useful to summarize the steps to follow in obtaining the generalized
results for any value of $J$.

Given any value of $J$, the first order of business is to label the $2J+1$
projections so that, starting from the maximum projection, the index $1$
corresponds to the projection $J\hbar $, the index $2$ to the projection $%
(J-1)\hbar $, and so on until the index $2J+1$ which corresponds to the
projection $-J\hbar $. Next we obtain the standard form of the operator $%
[\sigma _z].$ In units of $\hbar $, this will be

\begin{equation}
\lbrack \sigma _z]=\left( 
\begin{array}{ccccc}
J & 0 & ... & 0 & 0 \\ 
0 & J-1 & ... & 0 & 0 \\ 
... & ... & ... & ... & ... \\ 
0 & 0 & ... & -J+1 & 0 \\ 
0 & 0 & ... & 0 & -J
\end{array}
\right) .  \label{hu199}
\end{equation}
The normalized vectors of this operator will be

\begin{equation}
\lbrack \xi _J]=\left( 
\begin{array}{c}
1 \\ 
0 \\ 
... \\ 
0
\end{array}
\right) ,\;[\xi _{J-1}]=\left( 
\begin{array}{c}
0 \\ 
1 \\ 
... \\ 
0
\end{array}
\right) ,......,[\xi _{-J}]=\left( 
\begin{array}{c}
0 \\ 
0 \\ 
... \\ 
1
\end{array}
\right) ,\;\text{etc}.  \label{tw200}
\end{equation}
corresponding to the eigenvalues $J$, $(J-1)$,....,$-J$ respectively.

Next we deduce the ladder operators by means of their action on the
eigenvectors of the $z$ component. From them we get the $x$ and $y$
components of the spin. Using the three components of the spin, we obtain
the operator 
\[
\lbrack {\bf \sigma }\cdot \widehat{{\bf a}}]=[\sigma _{\widehat{{\bf a}}}]=(%
\widehat{{\bf i}}[\sigma _x]+\widehat{{\bf j}}[\sigma _y]+\widehat{{\bf k}}[%
\sigma _z])\cdot \widehat{{\bf a}} 
\]
where $\widehat{{\bf a}}$ $=(\sin \theta \cos \varphi ,\sin \theta \sin
\varphi ,\cos \theta )$.

The eigenvectors of this operator will contain as their elements probability
amplitudes of the form $\chi (m_i^{(\widehat{{\bf a}})};m_f^{(\widehat{{\bf d%
}})})$, where $\widehat{{\bf d}}$ is a vector which has to be deduced$.$
Similarly, if we now seek the eigenvectors of the operator $[{\bf \sigma }%
\cdot \widehat{{\bf c}}]$, where $\widehat{{\bf c}}=(\sin \theta ^{\prime
}\cos \varphi ^{\prime },\sin \theta ^{\prime }\sin \varphi ^{\prime },\cos
\theta ^{\prime }),$ we obtain all the probability amplitudes $\chi (m_i^{(%
\widehat{{\bf c}})};m_f^{(\widehat{{\bf d}})})$ because they will be the
elements of the eigenvectors of this operator. We then use the Land\'e
expansion Eqn. (\ref{two}), to eliminate $\widehat{{\bf d}}$ in order to
obtain the generalized probability amplitudes which represent spin
projection measurements from $\widehat{{\bf a}}$ to $\widehat{{\bf c}}$.

Armed with the generalized probability amplitudes, we obtain the generalized
operator for the $z$ component of the spin. We have to keep in mind that
these are defined in terms of an intermediate reference vector $\widehat{%
{\bf b}}$, and the final reference direction $\widehat{{\bf c}}$ with
respect to which we seek the spin projection. Let the spin projections with
respect to the vector $\widehat{{\bf b}}$ be labelled from the largest to
the smallest as $b_1$, $b_2$,....$b_{2J+1}.$ Let the spin projections with
respect to the vector $\widehat{{\bf c}}$ be labelled in the same way. Then
the elements of $[\sigma _{\widehat{{\bf c}}}]$ are

\[
(\sigma _{\widehat{{\bf c}}})_{ij}=\sum\limits_{k=1}^{2J+1}\phi
(b_j;c_k)\phi (c_k;b_i)r_k. 
\]

In this formula, $r_k$ is the $k$th projection of the spin with respect to
the direction $\widehat{{\bf c}}$, measured in units of $\hbar $.

The eigenvectors of the generalized operator $[\sigma _{\widehat{{\bf c}}}]$
contain as their elements generalized probability amplitudes. In order for
them to be eigenvectors of $[\sigma _{\widehat{{\bf c}}}],$ the initial
direction in the probability amplitudes must be $\widehat{{\bf c}}$, while
the final must be $\widehat{{\bf b}}$. We arrange the amplitudes into column
vectors with all those starting from one projection belonging together. With
regard to the final direction, the probability amplitudes are arranged so
that the one corresponding to the maximum value of the projection comes
first, then the one corresponding to the next largest projection and so on.

With both $[\sigma _{\widehat{{\bf c}}}]$ and its eigenvectors known, we
obtain $[\sigma _x]$ and its eigenvectors by applying the transformation of
the kind $\varphi ^{\prime }\rightarrow \varphi ^{\prime }+\pi /2$ or $%
\varphi ^{\prime }\rightarrow \varphi ^{\prime }-\pi /2$ on these
quantities. It would appear from the two cases so far seen that in general,
a little trial and error might be necessary. We have selected the correct
transformation if when we set $\theta =\theta ^{\prime }$ and $\varphi
=\varphi ^{\prime }$ we obtain the standard results.

Transformations of the form $\theta ^{\prime }=\pi /2$ and $\varphi =\varphi
^{\prime }+\pi /2$ or $\varphi =\varphi ^{\prime }-\pi /2$ will give us $%
[\sigma _y]$ and its eigenvectors; again the requirement that the standard
forms be obtained in the limit $\theta =\theta ^{\prime }$ and $\varphi
=\varphi ^{\prime }$ will determine which of these transformations is
correct. With all three components known, the raising and lowering operators
can also be obtained.

\section{\bf Discussion}

In this paper we have extended the ideas we first introduced in ref. $1$ by
using them to derive generalized operators and vectors as well as
probability amplitudes for spin-$1$ systems. That we have succeeded in doing
so lends strong support to the ideas of Land\'e, which underlie our work. We
consider it quite remarkable and important that the matrix treatment of spin
can be derived from a probability amplitude basis. It should be remembered
that spin is normally represented as being fundamentally of matrix nature.
But here we have given spin a treatment analogous to that given to orbital
angular momentum whose matrix description is based on the spherical
harmonics. This shows that in the treatment of spin, matrices can be
dispensed with if necessary. Perhaps other quantities in quantum mechanics
that are treated by a purely matrix approach can also be reduced to a
probability amplitude basis. The case of isotopic spin comes to mind.

Our method depends crucially on the idea that a probability amplitude must
always contain two state labels, one label being for the state that obtains
before measurement and the other for the state that results from
measurement. If we accept that the main difference between the spin
probability amplitude and the ordinary wave function is that the former
refers to discrete final states while the latter refers to continuous final
states, then we may interpret the wave function differently from the
standard way. Since the wave function in coordinate space results as the
solution of an energy eigenvalue equation which is cast in terms of the
coordinates, it is seen to be the probability amplitude for obtaining
various values of position if the measurement starts from a state
characterized by the corresponding energy eigenvalue. Likewise, for a given
energy eigenvalue, the wave function in momentum space is the probability
amplitude for obtaining various values of the momentum upon measurement if
the initial state is characterized by the energy eigenvalue. For a system
with a spherically symmetric potential, the radial solution is the
probability amplitude for obtaining various values of the radial coordinate
if the measurement starts from a state defined by the energy eigenvalue. The
spherical harmonics are probability amplitudes for obtaining various angular
positions starting from a state characterized by the angular momentum
corresponding to the values of $l$ and $m$.

We notice that the results we have presented in this paper also apply to $%
l=1 $ orbital angular momentum. They obtain if we measure the angular
momentum projection twice. This kind of measurement is seen to be different
from that which the spherical harmonic represents. As stated already, a
spherical harmonic connects an angular momentum state and a position state,
while the probability amplitudes presented in this paper connect two angular
momentum states or two spin states.

It is the practice to give the standard forms of the spin vectors when
constructing a complete wave function for a quantum system. We now realize
that in order for the limited standard forms to be useful, we must be
careful to use the generalized operators when actually computing such
quantities as expectation values. We must match spin vectors and operators
properly in order to obtain the correct results. The intermediate reference
direction connects the quantities in the expressions for the vectors with
the quantities in the expressions for the operator. The final direction in
the expressions for the vectors must correspond to the initial direction in
the expressions for the elements of the operator. In this connection, we
should distinguish clearly between the vectors for general calculations of
the expectation value and those which are eigenvectors of the operator. The
former refer to an initial reference direction which is different from the
final direction with respect to which we seek the spin projection. The
latter on the contrary are such that the initial direction equals the final
direction. Thus the vectors that we must use if we seek to calculate the
expectation value when the spin measurement is a repeat measurement are the
eigenvectors of the operator.

We have seen that matrix mechanics vectors do not directly give the
probabilities of finding eigenvalues of interest in individual measurements.
According to the Land\'e view, this is however what the wave function gives.
Nevertheless, the vectors are extremely compact because each vector does
contain all the probability amplitudes.

In conclusion, we have obtained from first principles new generalized
formulas for spin $1$ operators and vector states by means of the Land\'e
interpretation of quantum mechanics. It is our belief that a re-examination
of other standard quantum results in the light of this approach will yield
useful generalizations of these results and provide valuable insights into
quantum theory.

\section{\bf References}

1. Mweene H. V., ''Derivation of Spin Vectors and Operators From First
Principles'' submitted to Foundations of Physics, quant-ph/9905012

2. Mweene H. V., ''Generalized Spin-1/2 Operators and Their Eigenvectors'',
quant-ph/9906043

3. Land\'e A., ''New Foundations of Quantum Mechanics'', Cambridge
University Press, 1965.

4. Land\'e A., ''From Dualism To Unity in Quantum Physics'', Cambridge
University Press, 1960.

5. Land\'e A., ''Foundations of Quantum Theory,'' Yale University Press,
1955.

6. Land\'e A., ''Quantum Mechanics in a New Key,'' Exposition Press, 1973.

\end{document}